\documentclass[galaxies,review,accept,oneauthor]{mdpi}

\usepackage{booktabs}
\usepackage{multirow}
\usepackage{soul}
\usepackage{microtype}
\firstpage{1}
\pubvolume{00}
\issuenum{0}
\articlenumber{0}
\pubyear{2019}
\copyrightyear{2019}
\history{Received: 8 September 2019; Accepted: 4 November 2019; Published: date}

\Title{Relativistic Jets in Gamma-Ray-Emitting Narrow-Line Seyfert 1 Galaxies}

\Author{Filippo D'Ammando}

\address[1]{%
INAF---Istituto di Radioastronomia, Via P. Gobetti 101, I-40129 Bologna, Italy; dammando@ira.inaf.it}

\abstract{Before the launch of the {\em Fermi Gamma-ray Space Telescope} satellite only two classes of active galactic nuclei  (AGN) were known to generate relativistic jets and thus to emit up to the $\gamma$-ray energy range: blazars and radio galaxies, both hosted in giant elliptical galaxies. The discovery by the Large Area Telescope (LAT) on-board the {\em Fermi} satellite of variable $\gamma$-ray emission from a few radio-loud narrow-line Seyfert 1 galaxies (NLSy1) revealed the presence of an emerging third class of AGN with powerful relativistic jets. Considering that NLSy1 are usually hosted in late-type galaxies with relatively small black hole masses, this finding opened new challenging questions about the nature of these objects, the disc/jet connection, the emission mechanisms at high energies, and the formation of relativistic jets.
In this review, I will discuss the broad-band properties of the $\gamma$-ray-emitting NLSy1 included in the Fourth {\em Fermi} LAT source catalog, highlighting major findings and open questions regarding jet physics, black hole mass estimation, host galaxy and accretion process of these sources in the {\em Fermi}~era.}


\begin{document}

\section{Introduction}

Only a small percentage of active galactic nuclei (AGN) is radio-loud, and~this characteristic is commonly ascribed to the presence of a relativistic jet, roughly perpendicular to the accretion disc. Relativistic jets are the most extreme manifestation of the power that can be generated by a super-massive black hole (SMBH) in the center of an AGN. Their emission is observed across the entire electromagnetic spectrum with a large fraction of the power emitted in $\gamma$-rays. Before~the launch of the {\em Fermi Gamma-ray Space Telescope} satellite only two classes of AGN were known to generate strong relativistic jets, therefore to emit up to the $\gamma$-ray energy range: blazars and radio galaxies \citep{hartman99}, both~hosted in giant elliptical galaxies \citep{blandford78}. The~observations by the Large Area Telescope (LAT) on board the {\em Fermi} satellite confirmed that the extragalactic $\gamma$-ray sky is dominated by blazars, with~only a few radio galaxies detected \citep{acero15,abdollahi19}. However, the~discovery by {\em Fermi}-LAT of variable $\gamma$-ray emission from a few radio-loud narrow-line Seyfert 1 galaxies (NLSy1) revealed the presence of a possible third class of AGN with powerful relativistic jets\footnote{Based on the new classification proposed by \citep{padovani17}, these sources can be defined as ``jetted AGN''.} (e.g., \citep[]{abdo2009a,abdo2009b,dammando16a}).

NLSy1 represent a rare type of Seyfert galaxies identified by \citep{osterbrock85} and characterized by their optical properties: narrow permitted emission lines (full width at half maximum, FWHM (H$\beta$)~$<$~2000~km~s$^{-1}$), weak [OIII]/$\lambda$5007 emission lines (i.e.,~[OIII]/H$\beta$ $<$ 3), and~usually strong Fe II emission complexes (e.g.,~\citep[]{goodrich89,pogge00}). In~X-rays they exhibit strong and rapid variability, steep~spectra, relatively high luminosity, and~substantial soft X-ray excess below 2 keV (e.g.,~\citep[]{boller96, leighly99, grupe10}).~These~observational characteristics point to systems with smaller masses of the central SMBH (10$^6$--10$^8$~M$_\odot$) and higher Eddington ratios (close to or above the Eddington limit) with respect to blazars and radio galaxies (e.g.,~\citep[]{zhou06, yuan08, rakshit17}). Optical estimates of BH mass suggested that NLSy1 lies below the M$_{BH}$--$\sigma_{\rm\,bulge}$ relation, where $\sigma_{\rm\,bulge}$ is the velocity dispersion of the galaxy bulge (e.g.,~\citep[]{ferrarese00, gebhardt00}). This suggested that NLSy1 are accreting at very high rate (e.g., \citep[]{grupe04}) and are objects in the early stages of their evolution (e.g.,~\citep[]{mathur01}). However, the~use of optical methods to determine the BH mass of NLSy1 has been challenged by several authors. In~fact, the~effects of radiation pressure on the broad line region (BLR) clouds should be higher in case of highly accreting AGN, leading to underestimated masses~\citep{marconi08}. In~the same way, in~case of a disc-like geometry of the BLR, projection effects could explain the smaller masses estimated in NLSy1 \citep{decarli08}. Fitting the optical spectrum of a large sample of NLSy1, \citep{viswanath19} found a mean BH mass of $\sim$10$^{8}$ M$_\odot$ for both radio-quiet and radio-loud NLSy1 similar to the values obtained for broad-line Seyfert galaxies. The~masses derived by the accretion disc model are one order of magnitude larger than their virial estimates. Virial estimates are prone to large uncertainties (e.g., \citep[]{mejia18}), suggesting that values obtained by fitting the accretion disc model, independent of the geometry and the kinematics of the BLR, should be more realistic (e.g., \citep[]{capellupo15}).

The huge spectroscopic database of the Sloan Digital Sky Survey (SDSS; \citep[]{york00}) enables the selection of large NLSy1 samples (e.g., \citep[]{zhou06,yuan08,rakshit17}). NLSy1 are generally radio-quiet (radio-loudness $R$ being defined here as ratio of rest-frame 1.4 GHz and 4400 \AA\, flux densities, $R$ $>$ 10), with~only a small fraction ($<$ 7$\%$; \citep[]{komossa06, zhou06, cracco16, rakshit17}) classified as radio-loud, while $\sim$15$\%$ of quasars are~radio-loud (e.g.,~\citep[]{kellerman16}). NLSy1 with higher values of radio-loudness ($R$ $>$ 100) are even more sparse ($\sim$3\%). Radio-loud NLSy1 usually show a compact morphology with a core-jet structure extending up to a few parsec, with~some cases in which the radio emission extends on kpc scales \citep{anton08, doi12, richards15, berton18, rakshit18}. Together with high values of brightness temperature and core dominance this suggested the presence of a relativistic jet (e.g., \citep[]{doi11}). In~the past several authors have investigated the peculiarities of radio-loud NLSy1 with non-simultaneous radio-to-X-ray data, suggesting similarities with the young stage of quasars or different types of blazars (e.g., \citep[]{komossa06,yuan08,foschini09}). The~detection of a few radio-loud NLSy1 in $\gamma$-rays by {\em Fermi}-LAT has been an unchallengeable evidence of the presence of highly relativistic jets in this class of AGN. In~addition, apparent superluminal jet components were observed in SBS 0846$+$513 \citep{dammando13a}, PMN J0948$+$0022 and 1H 0323$+$342 \citep{lister16}, suggesting high ($>$10) Lorentz factors and small ($<$10$^\circ$) viewing~angles.

NLSy1 are usually hosted in spiral galaxies (e.g., \citep[]{krongold01, deo06, ohta07, orban11, mathur12}), although~some objects are associated with early type S0 galaxies (e.g., Mrk 705 and Mrk 1239; \citep[]{markarian89}). The~presence of a relativistic jet in these sources seems to be in contrast to the paradigm that powerful jets could be produced only in elliptical galaxies (e.g., \citep[]{bottcher02,marscher09}). Indeed, the~most powerful jets are found in luminous elliptical galaxies with very massive (10$^8$--10$^{10}$ M$_\odot$) SMBH (e.g., \citep[]{sikora07}). This is interpreted as indirect evidence that a high spin is required for the jet production. In~fact, elliptical galaxies should be the result of a major merger event that is required for spinning up the central BH (e.g., \citep[]{sikora09}). The~detection of variable $\gamma$-ray emission from radio-loud NLSy1 poses intriguing questions about the nature of these sources, the~production of relativistic jets, and~the mechanisms of high-energy emission in the different class of AGN. In~this context the study of NLSy1 has received increasing~attention.

Nine NLSy1 are included in the recent Fourth {\em Fermi} LAT source catalog (4FGL, \citep[]{abdollahi19}), which covers 8 years (i.e., 4 August 2008--2 August 2016) of LAT data in the energy range from 50 MeV to 1 TeV. In~this paper I will review the radio-to-$\gamma$-ray properties of the $\gamma$-ray-emitting sources present in the 4FGL catalog and classified as bona-fide NLSy1. Although~biased by the author’s scientific interests, a~balanced review of relevant works on the selected topics is attempted. The~paper is organized as follows. The $\gamma$-ray properties of NLSy1 are discussed in Section~\ref{LAT_gamma}, while Sections~\ref{Xray_properties}--\ref{sec5} are devoted to the X-ray, infrared-optical, and~radio properties of the $\gamma$-ray-emitting NLSy1, respectively. Results about the spectral energy distribution (SED) modelling of these sources are reported in Section~\ref{SED}, while their BH mass and host galaxy are discussed in Section~\ref{host_mass}. Finally, I~draw some conclusions in Section~\ref{sec8}. Throughout the paper, a~$\Lambda$ cold dark matter cosmology with $H_0$ = 67 km s$^{-1}$ Mpc$^{-1}$, $\Omega_{\Lambda} = 0.68$ and $\Omega_{\rm m} = 0.32$ \citep{ade16} is~adopted.

\section{The $\gamma$-ray~Properties}\label{LAT_gamma}

Four radio-loud NLSy1 galaxies have been detected at high significance by {\em Fermi}-LAT in the first year of operation (i.e.,~1H 0323$+$342, PMN J0948$+$0022, PKS 1502$+$036, and~PKS 2004$-$447\footnote{PKS 2004$-$447 has a weak Fe II emission line with respect to the typical NLSy1 (for a discussion about the nature of this source see e.g., \citep[]{oshlack01,gallo06, baldi16}).}) and are included in the First {\em Fermi} LAT source catalog, based on the first 11 months of LAT operation (1FGL;~\citep[]{abdo10}). No new $\gamma$-ray emitting NLSy1 has been reported in the Second {\em Fermi} LAT source catalog, including observations performed between 4 August 2008 and 31 July 2010 (2FGL; \citep[]{nolan12}), with~respect to the 1FGL catalog. SBS 0846$+$513 was detected in $\gamma$-rays for the first time during October 2010--August 2011, when a significant increase of activity was observed by LAT~\citep{dammando12}. Thus in the Third {\em Fermi} LAT source catalog (3FGL), based on the first 4 years of LAT operation, five NLSy1 have been reported~\citep{acero15}.
The~first $\gamma$-ray detection of FBQS J1644$+$2619 has been reported in \citep{dammando15a}, analysing the $\gamma$-ray data collected on a longer period than that of 3FGL, i.e.,~ August 2008--December 2014. Both the LAT detection of high $\gamma$-ray activity periods from FBQS J1644$+$2619 observed from November 2008 to January 2009 and July to October 2012 correspond to periods of high optical activity, as~observed in $V$-band by the Catalina Real-Time Transient survey. That has confirmed the association of the $\gamma$-ray source with the NLSy1 FBQS J1644$+$2619. With~the new Pass 8 event-level analysis (e.g., \citep[]{atwood13}) and by using 7~and 8.5 years of {\em Fermi}-LAT data other two new NLSy1 have been detected for the first time in $\gamma$-rays, B3 1441$+$476 and TXS 2116$-$077, as~reported in \citep{dammando16a,paliya18}, respectively. Another $\gamma$-ray-emitting NLSy1 has been included in the Fourth {\em Fermi} LAT source catalog (4FGL, \citep[]{abdollahi19}), IERS B1303$+$515, for~a total of nine NLSy1 reported in the 4FGL~catalog. 

The $\gamma$-ray properties of the radio-loud NLSy1 included in the 4FGL catalog are reported in Table~\ref{LAT_properties}. The~nine NLSy1 detected at high significance by {\em Fermi}-LAT up to now span a redshift range between 0.061 and 0.787. Luminosity, variability and spectral properties of the NLSy1 in $\gamma$-rays indicate a blazar-like behaviour (e.g., \citep[]{dammando16a,paliya18}). The~average apparent $\gamma$-ray isotropic luminosity of these sources in the 0.1--300 GeV energy band is between 2 $\times$ 10$^{44}$ and 8 $\times$ 10$^{46}$ erg s$^{-1}$, a~range of values typical of blazars, in~particular of FSRQ. This is an indication of a relatively small viewing angle with respect to the jet axis for these NLSy1, and~therefore a high beaming factor for their $\gamma$-ray emission, similarly to what is inferred for~FSRQ.

\begin{table}[H]
\centering
\caption{The $\gamma$-ray properties of radio-loud NLSy1 detected by {\em Fermi}-LAT. Energy flux and spectral parameters are taken from the 4FGL catalog \citep{abdollahi19}.}
\label{LAT_properties}
\scalebox{.94}[.94]{\begin{tabular}{ccccccc}
\toprule
\textbf{Source Name} & \textbf{Redshift} & \textbf{Energy Flux (E \boldmath{$>$} 0.1 GeV)} & \boldmath$\Gamma_\gamma$ & \boldmath$\alpha$ & \boldmath$\beta$ & \boldmath{\textbf{L}$_{\gamma}$} \\
& \boldmath$z$ & \boldmath$\times 10^{-11}$ \textbf{erg cm}$^{-2}$ \textbf{s}$^{-1}$ & & & & \boldmath{\textbf{erg s}$^{-1}$} \\
\midrule
1H 0323$+$342       &  0.061  & $2.17 \pm 0.10$  & $2.82\pm0.04$ & $2.72\pm0.05$  & $0.10\pm0.03$ & 2.1 $\times$ 10$^{44}$ \\
SBS 0846$+$513      &  0.584  & $2.17 \pm 0.08$  & $2.17\pm0.02$ & $2.27\pm0.04$  & $0.08\pm0.02$ & 3.2 $\times$ 10$^{46}$  \\
PMN J0948$+$0022    &  0.585  & $5.00 \pm 0.10$  & $2.64\pm0.02$ & $2.46\pm0.03$  & $0.16\pm0.02$ & 7.5 $\times$ 10$^{46}$  \\
IERS B1303$+$515    &  0.787  & $0.22 \pm 0.04$  & $2.85\pm0.17$ & --             & --            & 6.9 $\times$ 10$^{45}$  \\
B3 1441$+$476       &  0.705  & $0.20 \pm 0.04$  & $2.56\pm0.11$ & $2.25\pm0.26$  & $0.41\pm0.19$ & 4.7 $\times$ 10$^{45}$  \\
PKS 1502$+$036      &  0.408  & $1.61 \pm 0.08$  & $2.59\pm0.04$ & $2.48\pm0.06$  & $0.10\pm0.03$ & 1.0 $\times$ 10$^{46}$  \\
FBQS J1644$+$2619   &  0.145  & $0.45 \pm 0.06$  & $2.78\pm0.10$ & --             & --            & 2.7 $\times$ 10$^{44}$  \\
PKS 2004$-$447      &  0.240  & $0.92 \pm 0.08$  & $2.60\pm0.05$ & $2.41\pm0.09$  & $0.18\pm0.06$ & 1.7 $\times$ 10$^{45}$  \\
TXS 2116$-$077      &  0.260  & $0.32 \pm 0.06$  & $2.83\pm0.15$ & --             & --            & 7.2 $\times$ 10$^{44}$  \\
\bottomrule
\end{tabular}}
\end{table}

In the same way, the~average photon index $\Gamma_\gamma$ of the $\gamma$-ray-emitting NLSy1 ranges between 2.2 and 2.9, a~range of values usually observed in FSRQ \citep{ackermann15, abdollahi19}. For~six of the nine sources the $\gamma$-ray spectrum is significantly curved ($TS_{curv}$ $>$ 9 \footnote{$TS_{curv}$ = 2 log(Likelihood(log-parabola)/Likelihood(power-law)), i.e.,~a comparison of the likelihood function changing only the spectral representation of the source.}), and~a log-parabola model is preferred over a simple power-law model. In~that case the spectral slope ($\alpha$) and  curvature parameter ($\beta$) have been reported in Table~\ref{LAT_properties}, together with the photon index obtained with the power-law model. A~significant spectral curvature is usually observed in the $\gamma$-ray spectra of FSRQ, suggesting similarities in the physical processes at work in these two classes of $\gamma$-ray-emitting AGN. The~jet powers of $\gamma$-ray-emitting NLSy1 (obtained by the SED modelling) are similar to the values obtained for low-synchrotron-peaked BL Lacs and at the low end of the FSRQ distribution (e.g., \citep[]{foschini11, dammando13a, paliya14, dammando15b, dammando16b}). However, the~jet power depends on the BH mass, therefore the use of an underestimated BH mass value (see Section~\ref{host_mass}) leads to an underestimated jet power. After~correcting for higher SMBH values, the~jet powers of $\gamma$-ray-emitting NLSy1 should be comparable to those usually obtained for~FSRQ.

\begin{figure}[H]
\centering
\rotatebox{0}{\resizebox{!}{70mm}{\includegraphics{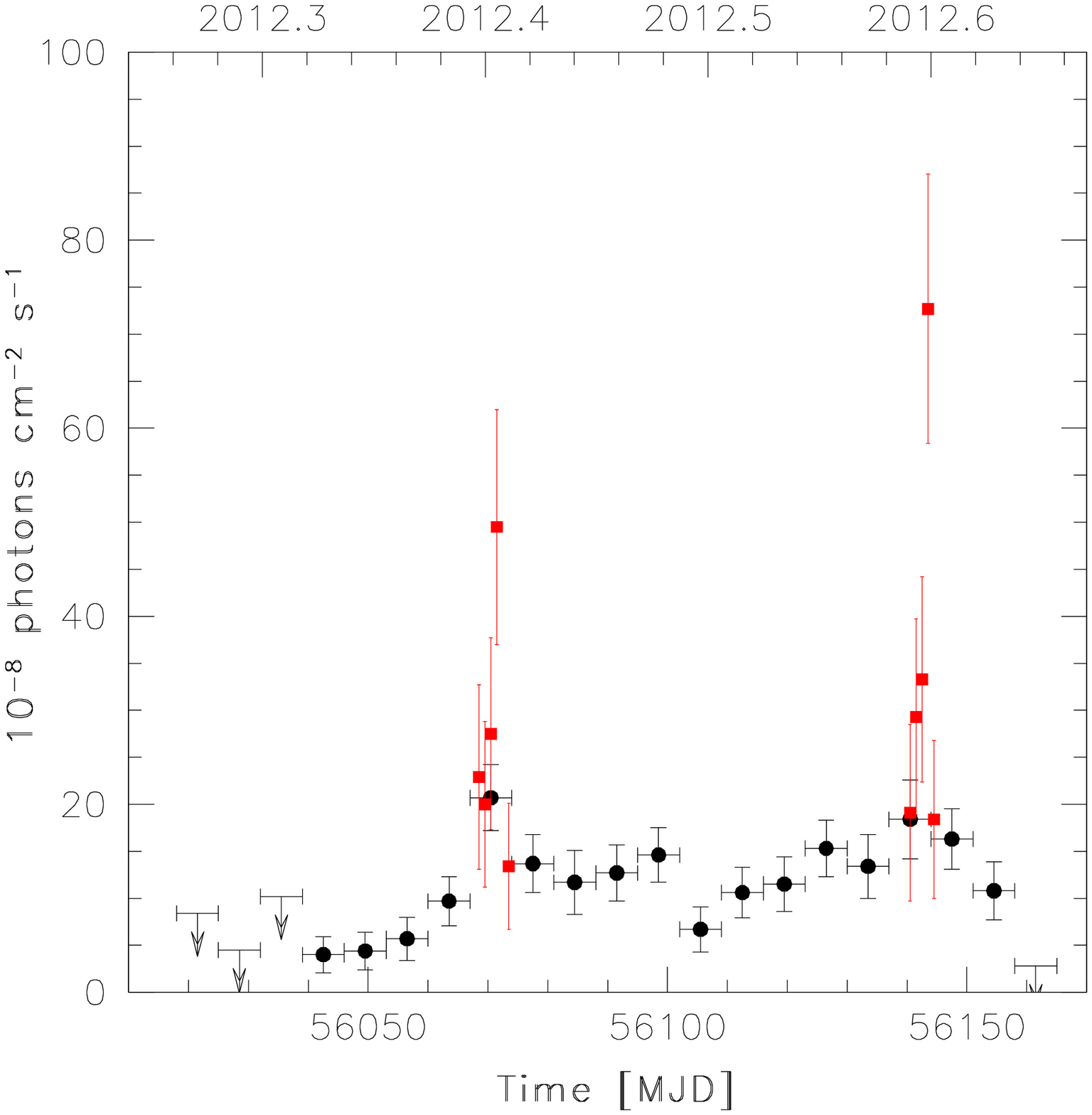}}}
\hspace{0.1cm}
\rotatebox{0}{\resizebox{!}{70mm}{\includegraphics{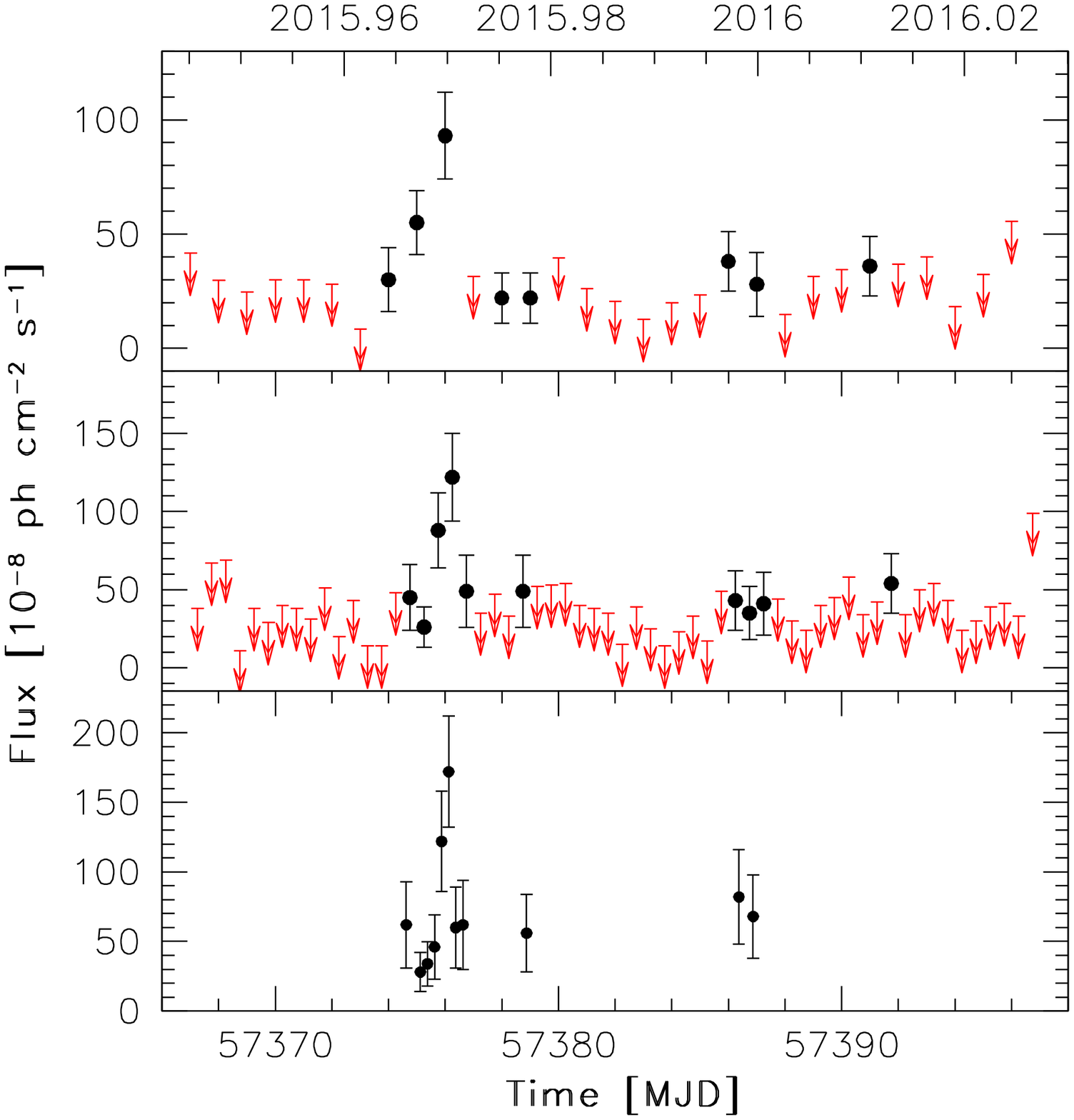}}}
\caption{{\it Left panel}: LAT light curve of SBS 0846$+$513 in the 0.1--100 GeV energy range during 1 April--28 August 2012 with 7 days or 1 day (shown as red squares) time bins. Adapted from \citep{dammando13a}. {\it Right~panel}: LAT light curve of PKS 1502$+$036 in the 0.1--300 GeV energy range during 11 December 2015--9 January 2016, with~1-d time bins (top panel), 12-h time bins (middle panel), and~6-h time bins (bottom~panel). Adapted from \citep{dammando16b}. In~both panels arrow refers to 2-$\sigma$ upper limits. Upper limits are computed when $TS$ $<$ 10.}
\label{LATflare}
\end{figure}

\begin{figure}[H]
\centering
\rotatebox{0}{\resizebox{!}{70mm}{\includegraphics{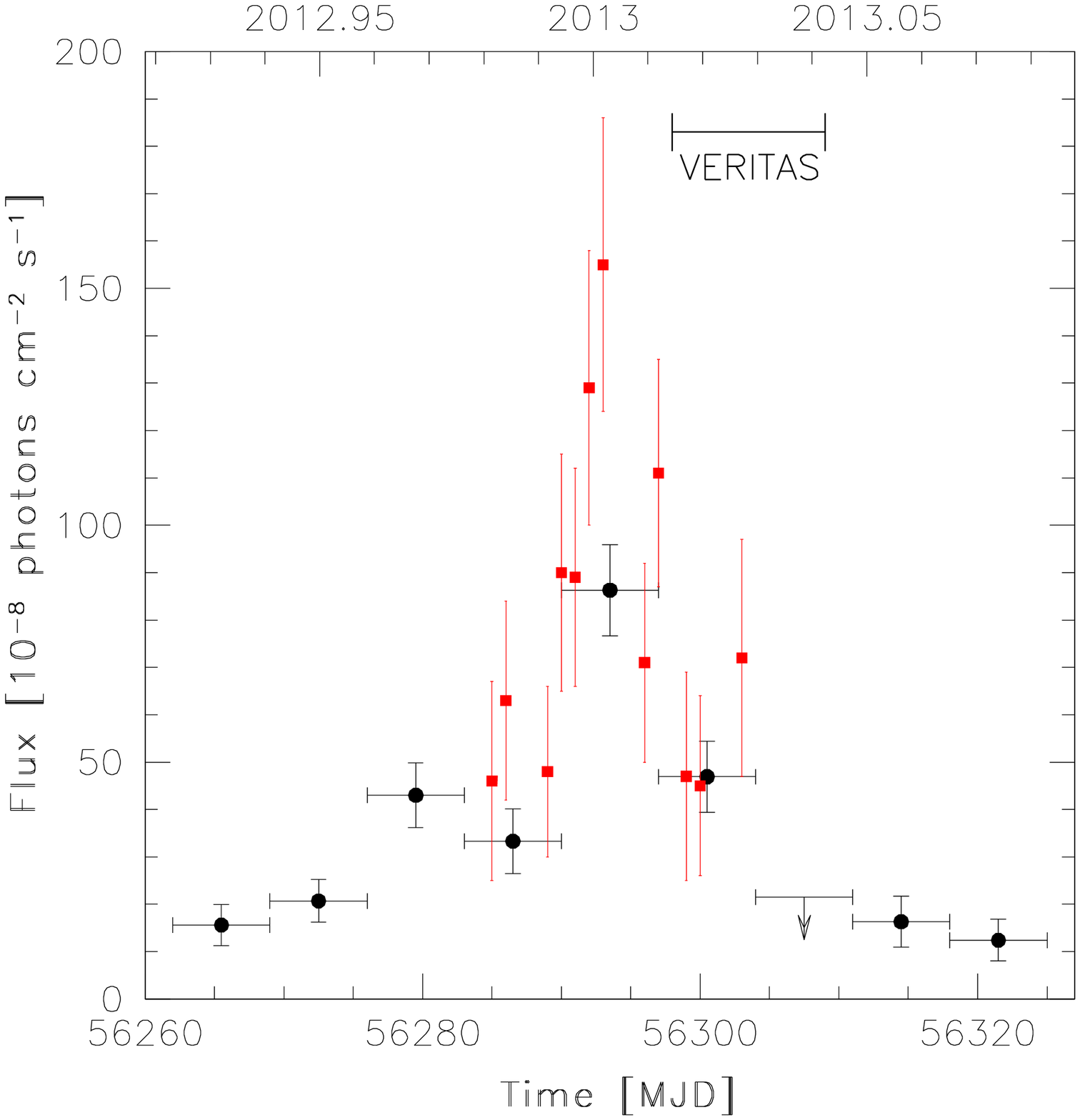}}}
\hspace{0.1cm}
\rotatebox{0}{\resizebox{!}{65mm}{\includegraphics{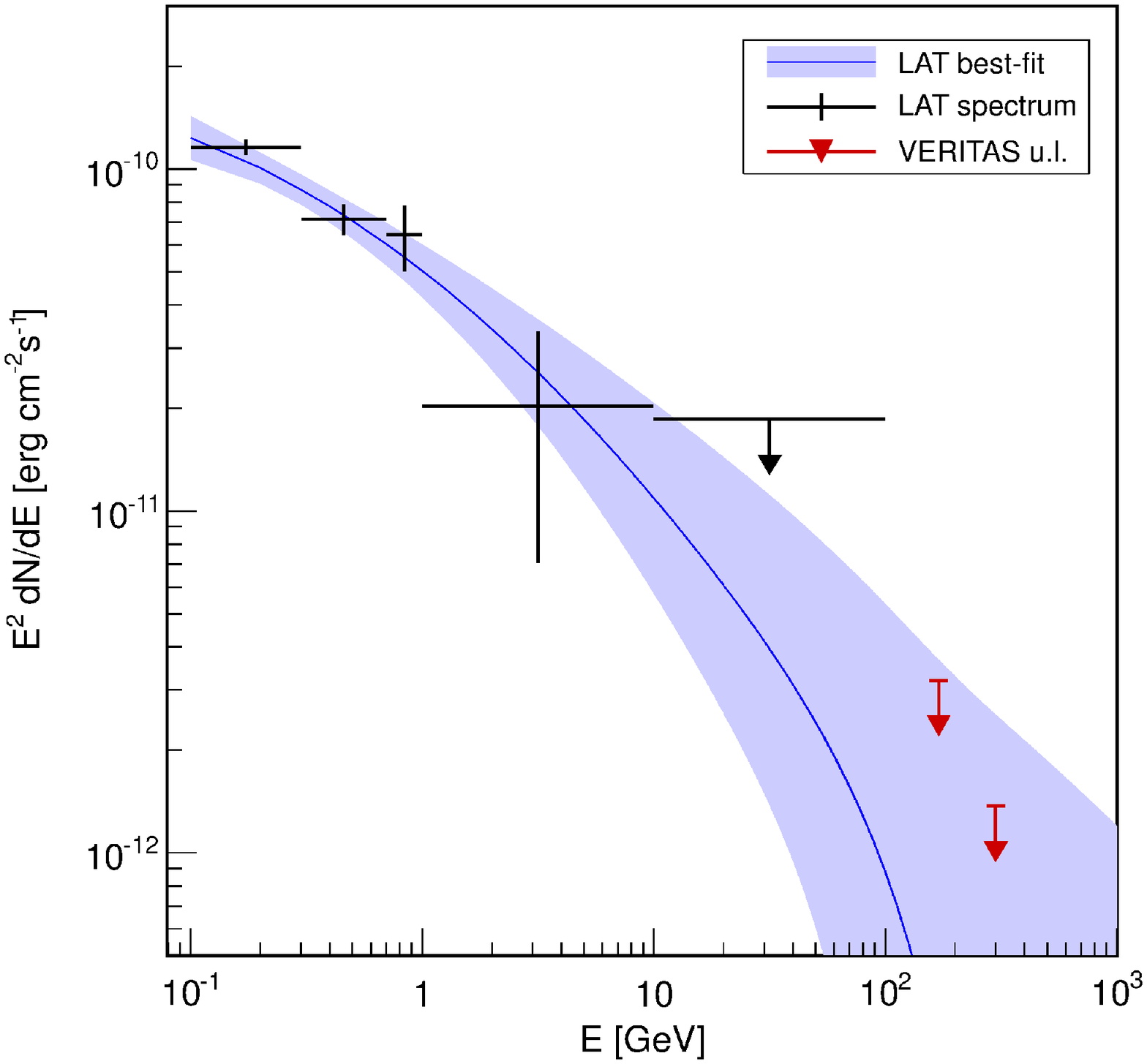}}}
\caption{{\it Left panel}: LAT light curve of PMN J0948$+$0022 in the 0.1--100 GeV energy range during 1 December 2012-- 31 January 2013, with 7-day time bins. Red solid squares represent daily fluxes reported for the period of high activity. The~horizontal line indicates the period of the VERITAS observation. Adapted from \citep{dammando15b}. {\it Right panel}: SED of PMN J0948$+$0022 in the MeV-to-TeV energy range. The~LAT spectrum was extrapolated to the TeV energies and corrected for EBL absorption using the model of~\citep{finke10}. {\em Fermi}-LAT and VERITAS data points and upper limits are shown. Adapted from \citep{dammando15b}. In~both panels, arrow refers to 2-$\sigma$ upper limits. Upper limits are computed when $TS$ $<$ 10.}
\label{LATflare2}
\end{figure}

Some of the $\gamma$-ray-emitting NLSy1 do exhibit large-amplitude $\gamma$-ray flares: SBS 0846$+$513 \citep{dammando12,dammando13a} (see Figure~\ref{LATflare}, left panel); PMN J0948$+$0022 \citep{dammando15b,foschini11b} (see Figure~\ref{LATflare2}, left panel),  1H 0323$+$342 \citep{paliya14}, PKS~1502$+$036 \citep{dammando16b}, PKS 2004$-$447 \citep{gokus19}, with an apparent isotropic $\gamma$-ray luminosity of $\sim$10$^{48}$ erg s$^{-1}$, comparable to that of bright FSRQ. This is another indication that NLSy1 are able to host relativistic jets as powerful as those in FSRQ. Variability was observed during all $\gamma$-ray flares on a daily time-scale, and~a subdaily time-scale in case of PKS 1502$+$036 \citep{dammando16b,paliya16a} (see Figure~\ref{LATflare},~right panel). The~$\gamma$-ray flaring activity of SBS 0846$+$513 and PMN J0948$+$0022 has been associated with a moderate spectral evolution, a~behaviour observed also in bright FSRQ and low-synchrotron-peaked BL Lacs (e.g., \citep[]{abdo10b}).

In addition to the nine NLSy1 reported in the 4FGL catalog, the~$\gamma$-ray detection of other NLSy1 has been claimed in literature: FBQS J1102$+$2239 \citep{foschini11}, FBQS J0956$+$2515 \citep{miller15}, NVSS J142106$+$385522~\citep{paliya18}, and~SDSS J164100.10$+$345452.7 \citep{lahteenmaki18}. However, these sources are not included in any LAT catalogue and their detection is not confirmed by dedicated analysis (e.g., \citep[]{dammando16a}). Moreover, SDSS J003159.85$+$093618.4 has been proposed as counterpart of the unidentified $\gamma$-ray source 3FGL J0031.6$+$0938 and classified as a NLSy1 \citep{paiano19}, but~the $\gamma$-ray source is not confirmed in the 4FGL catalog.
In addition, some confirmed $\gamma$-ray sources already present in the 3FGL catalogue has been tentatively classified as NLSy1 in \citep{paliya18}: 3FGL J0937.7$+$5008 (corresponding to GB6 J0937$+$5008), 3FGL J1520.3$+$4209 (corresponding to B3 1518$+$423), and~3FGL J2118.4$+$0013 (corresponding to PMN J2118+0013), based on the NLSy1 catalog prepared by \citep{rakshit17}. However, in~that catalog less restrictive criteria has been used to classify a source as a NLSy1 (i.e., FWHM(H$\beta$) $<$ 2200 km s$^{-1}$ and a strong Fe bump not needed) with respect to the historical ones. Although~no sharp change occurred at {FWHM values equal} to 2000 km s$^{-1}$, AGN properties change as a function of broad-line width, suggesting to keep a more conservative threshold as selection criterion. Moreover, in~some cases the determination of the FWHM of the emission line depends on the assumed line shape (Gaussian or Lorentzian) and the contribution of the narrow line component to the total line flux, as~in the case of 4C $+$04.42 (see~e.g.,~the~discussion in \citep[]{kynoch19}). Finally, the~uncertainties on the FWHM(H$\beta$) values of some sources reported in \citep{rakshit17} are quite large, making their classification uncertain. For~these reasons some of the sources included in that catalog can be considered just as candidate NLSy1 and not bona-fide NLSy1, including the three FSRQ tentatively re-classified by \citep{paliya18} and the $\gamma$-ray sources NVSS J093241$+$530633 and NVSS J095820$+$322401 reported in \citep{paliya18}\footnote{{In the same way, 3C 286 is classified as a Compact Steep Spectrum source in the 4FGL, while it is tentatively re-classified as a NLSy1 in \citep{berton17,liao19}. The~FWHM(H$\beta$) value reported in \citep{liao19} exceeds the historical threshold of 2000 km s$^{-1}$, therefore 3C 286 could be considered as a candidate NLSy1. It is worth nothing also that the inclination angle between the jet and our line of sight is 48 degrees \citep{an17}, significantly different from the relatively small angles estimated for the $\gamma$-ray-emitting NLSy1.}}. In~order to investigate the properties of the $\gamma$-ray-emitting NLSy1, it is important to start considering only bona-fide NLSy1, not~mixing these sources with FSRQ tentatively re-classified as NLSy1 and candidate NLSy1. For this reason this review is focused on the nine bona-fide NLSy1 included in the 4FGL catalog.

Among the 9 $\gamma$-ray-emitting radio-loud NLSy1, only one source, SBS 0846$+$513, is included in the Third catalogue of Hard {\em Fermi}-LAT sources (3FHL; \citep[]{ajello17}), based on 7 years of LAT data analysed in the 10 GeV--2 TeV energy range, with~a photon index of 2.93 $\pm$ 0.71 and a flux of (7.13 $\pm$ 1.50) $\times$ 10$^{-11}$ ph cm$^{-2}$ s$^{-1}$. There are no NLSy1 included in the Second {\em Fermi}-LAT catalog of High-Energy Sources (2FHL; \citep[]{ackermann16}), based on 80 months of data in the 50 GeV--2 TeV energy range. The~lack of $\gamma$-ray-emitting NLSy1 in the 2FHL is in agreement with the hard ($\Gamma_{\gamma}$ $>$ 2.2) $\gamma$-ray spectrum observed in the 0.1--300~GeV energy range over 8 years of {\em Fermi} operation (see Table~\ref{LAT_properties}) and the lack of detection at Very High Energy (VHE; E $>$ 100 GeV) by the current Cherenkov~Telescopes.

Pre-{\em Fermi} observations of 1H 0323$+$342 by Whipple \citep{falcone04} and VERITAS \citep{archambault16} at VHE resulted in upper limits. Furthermore, 1H 0323$+$342 and PKS 2004$-$447 were not detected by H.E.S.S. during an AGN monitoring program \citep{abramowski14}.
With the advent of {\em Fermi}-LAT, flaring activity observed in the High Energy (HE; 100 MeV $<$ 100 GeV) regime can trigger follow-up observations with the current generation of Imaging Atmospheric Cherenkov Telescope. VERITAS observed PMN J0948$+$0022 for $\sim$5.5 h during 6--13 January 2013 (see Figure~\ref{LATflare2}, left panel), a~few days after the $\gamma$-ray flare observed by {\em Fermi}-LAT on 1 January 2013, and~could set an upper limit of F$_{>\,0.2\rm\,TeV}$ $<$ 4 $\times$ 10$^{-12}$ ph~cm$^{-2}$~s$^{-1}$~\citep{dammando15b} (see Figure~\ref{LATflare2}, right panel). The~lack of detection of PMN J0948$+$0022 at VHE could be due to different reasons: (1) The distance of the source ($z$ = 0.5846) is relatively large and most of the GeV/TeV emission may be absorbed due to pair production from $\gamma$-ray photons of the source and the infrared photons from the extragalactic background light (EBL). However, it worth noting that at least three blazars, S3~0218$+$35 \citep{ahnen16}, PKS 1441$+$25 \citep{abeysekara15, ahnen15}, and~Ton 0599 \citep{mirzoyan17} are detected at VHE at much larger distance. (2)~Due to bad weather conditions, the~VERITAS observations were carried out a few days after the peak of the HE $\gamma$-ray activity, thus covering only the last part of the MeV-GeV flare. (3)~Considering the similarities with FSRQ, a~bright BLR may be present in the $\gamma$-ray-emitting NLSy1. The~presence of a BLR could produce a spectral absorption feature due to pair production, which may either suppress all $\gamma$-rays above a few GeV, or~manifest as a harder spectrum towards the VHE regime. The~geometry of the BLR does not affect the possible detection at VHE because the important parameter is the angle at which the jet sees the BLR (e.g., \citep[]{abdo2009a,tavecchio12,costamante18}). However, the~detection of FSRQ at VHE \citep{albert08, aleksic11a, aleksic14, aleksic11b, abramowski13, mirzoyan17, mukherjee17} has shown that the spectrum of FSRQ may extend to VHE energies during some flaring activities, indicating that the $\gamma$-rays may be produced outside the BLR during these high-activity periods. The~same scenario may apply to the $\gamma$-ray-emitting~NLSy1.

Future observations of $\gamma$-ray flaring NLSy1 observed by {\em Fermi}-LAT at VHE with the Cherenkov Telescope Array (CTA; \citep[]{CTA2019}) and other future $\gamma$-ray satellites  (e.g., AMEGO; \citep[]{meyer19}) will constrain the level of $\gamma$-ray emission of NLSy1 at 100 GeV or below. Preliminary simulations for follow-up observations of a sample of sources including bona-fide and candidate NLSy1 (not all of them with a confirmed detection in $\gamma$-rays by {\em Fermi}-LAT, see Section~\ref{LAT_gamma}) have shown that  SBS 0846$+$513, PMN J0948$+$0022, and~PKS 1502$+$036 should be detected with CTA during high activity states in 50 h of observations~\citep{romano18}.

\section{X-ray~Properties}\label{Xray_properties}

\begin{figure}[H]
\centering
\rotatebox{0}{\resizebox{!}{50mm}{\includegraphics{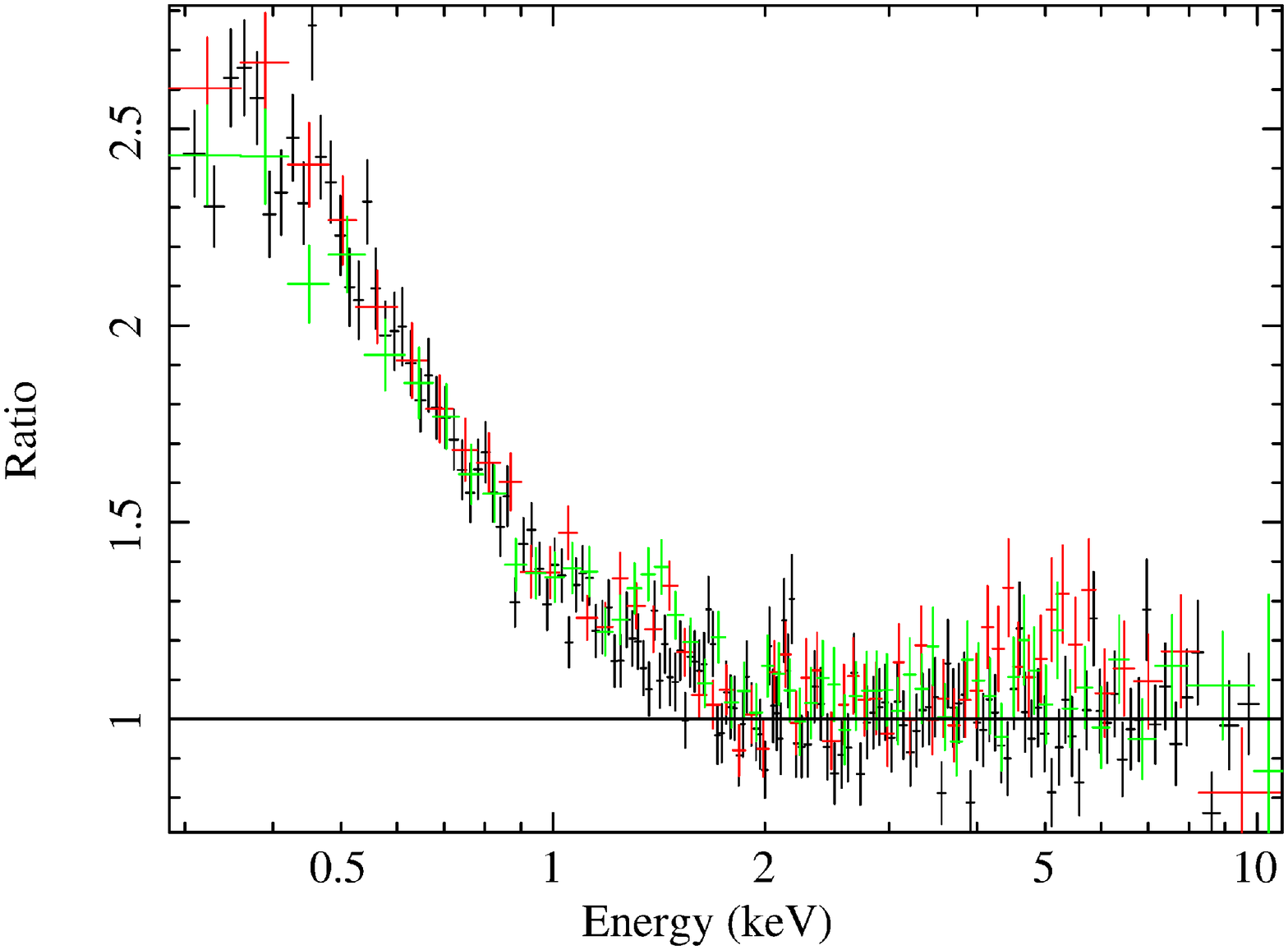}}}
\hspace{0.2cm}
\rotatebox{0}{\resizebox{!}{50mm}{\includegraphics{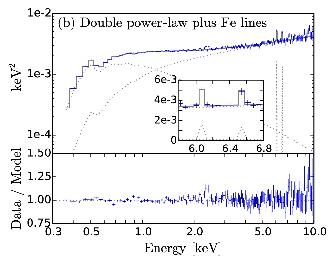}}}
\caption{{\it Left panel}: {\em XMM-Newton} EPIC pn (black), MOS1 (red), and~MOS2 (green) data of PMN J0948$+$0022 collected on 28--29 May 2011 shown as a ratio to a power law with photon index $\Gamma$ = 1.48. Adapted from \citep{dammando14}. {\it Right panel}: data/model ratio for the {\em XMM-Newton} EPIC pn X-ray spectrum of 1H 0323$+$342 collected on 23--24 August 2015. A~model including two power-laws and two Gaussian Iron emission lines with the Galactic value in the direction of the source fixed to N$_{\rm\,H,\,Gal}$ = 2.33 $\times$ 10$^{21}$ cm$^{-2}$ is applied to the data. Adapted from \citep{kynoch18}.}
\label{Xray}
\end{figure}

\begin{figure}[H]
\centering
\includegraphics[width=85mm]{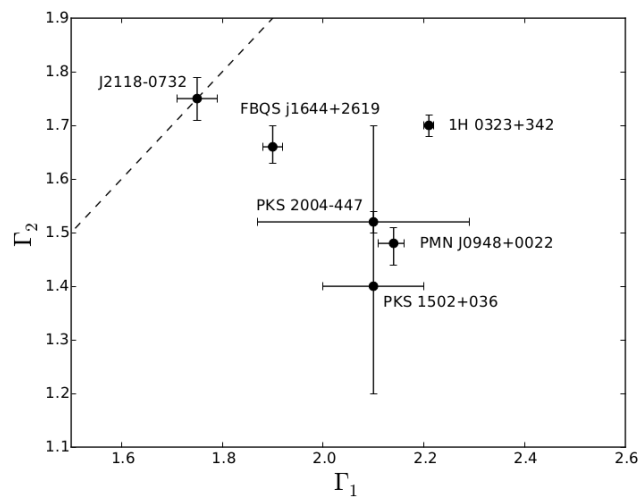}
\caption{Photon indices obtained from broken power-law fits to {\em XMM-Newton} spectra of $\gamma$-ray-emitting NLSy1. $\Gamma_{1}$ and $\Gamma_{2}$ are the photon indices below and above the break, respectively. One of the sources is best described by a simple power-law. In~case of 1H 0323$+$342 the broken power-law is just an approximation, as~the spectrum show additional complexity. Adapted from \citep{larsson18}.}
\label{Xray2}
\end{figure}

The X-ray spectra of radio-quiet NLSy1 are usually characterized by a steep photon index ($\Gamma_{\rm\,X}$~$>$~2,~\citep{grupe10, boller96}). The~X-ray emission is one of the most intriguing aspects of the $\gamma$-ray-emitting NLSy1, with~some differences observed with respect to FSRQ. The~spectra are hard above 2 keV ($\Gamma_{\rm\,X}$~$<$~2), which is similar to what is observed in FSRQ rather than radio-quiet NLSy1, suggesting that a significant contribution of inverse Compton (IC) radiation from a relativistic jet dominates that part of the X-ray spectrum. Differently from FSRQ, a~narrow Fe line at 6.4 keV has been reported for one source, 1H~0323$+$342 (Figure~\ref{Xray}, right panel), which has the softest X-ray spectrum of the $\gamma$-ray-emitting NLSy1~\citep{kynoch18, dammando17b}. The~majority of $\gamma$-ray-emitting NLSy1 for which good-quality X-ray spectra are available show {an excess of emission at low energies with respect to the extrapolation of the hard X-ray spectral continuum model}, making them different from typical blazars. As~an example, the~left panel of Figure~\ref{Xray} shows {this excess below 2 keV} in the X-ray spectrum of PMN J0948$+$0022. The~0.3--10~keV energy range spectra of $\gamma$-ray-emitting NLSy1 are usually well fitted by a broken power-law with a break around 2 keV \citep{larsson18}\footnote{In case of 1H 0323$+$342 a more complex X-ray spectrum was observed (e.g., \citep[]{kynoch18}).}, being the spectrum below the break due to disc/coronal emission, a~typical feature of Seyfert galaxies, and~the emission above the break dominated by the jet emission, a~typical feature of blazars. Photon indices obtained from power-law and broken power-law fits to the spectra of $\gamma$-ray-emitting NLSy1 observed with {\em XMM-Newton} are plotted in Figure~\ref{Xray2}. Only~two of these sources do not have a significant soft excess: TXS 2116$-$077~\citep{yang18} and PKS 2004$-$447~\citep{orienti15, kreikenbohm16}, although~a tentative soft excess was reported in one of the two {\em XMM-Newton} observations performed in 2012 for the latter source \citep{gallo06}. None of the sources { has shown} evidence of intrinsic absorption in~X-rays.

\unskip

A likely explanation for the X-ray spectra of $\gamma$-ray-emitting NLSy1 is that the underlying Seyfert emission, originating from the corona and accretion disc, has a noticeable contribution at low energies, the~so-called soft X-ray excess. The~origin of the soft X-ray excess is still an open issue both in radio-quiet and radio-loud AGN (e.g., \citep[]{gierlinski04}). Such a Seyfert component is a typical feature in the X-ray spectra of radio-quiet NLSy1, but~it is quite unusual in jet-dominated AGN, even if not unique (e.g.,~PKS 1510-089, \citep{kataoka08}; 3C 273, \citep{grandi04}; 4C $+$04.42  \citep{derosa08}). The~soft excess is often exceptionally strong in radio-quiet NLSy1, making it plausible that it would be detectable in the $\gamma$-ray-emitting radio-loud NLSy1 even though the jet emission is strong. In~agreement with this, the~analysis of the spectra of PMN J0948$+$0022 \citep{dammando14}, FBQS J1644$+$2619 \citep{larsson18}, and~1H 0323$+$342 \citep{kynoch18} can be well described by models that include reflection and/or Comptonisation in addition to the jet emission. PMN J0948$+$0022 and 1H 0323$+$342 {have shown} breaks in the root-mean square (RMS)-variability around the spectral breaks, lending further support to a scenario where different components dominate at low and high energies \citep{dammando14, kynoch18}. Another possibility is that the soft X-ray emission originates from the jet itself. This~could be the case if the tail of the synchrotron emission extends to the X-ray range, though~the SED modelling of these sources does not favour this scenario (see Section~\ref{SED}). It has also been suggested that bulk Comptonization by a blob of plasma travelling along the jet could produce an excess at soft X-rays \citep{celotti07}. A~similar feature has been proposed for the interpretation of the X-ray spectrum of the FSRQ PKS 1510$+$089 \citep{kataoka08} and the candidate NLSy1 4C $+$04.42 \citep{derosa08}. However, this feature should be transient, in~apparent contradiction with the fact that it is observed in the majority of sources. Currently, the~reason for the differences between the X-ray spectra of $\gamma$-ray-emitting NLSy1 and FSRQ is under~debate.

Further long-duration X-ray observations with {\em XMM-Newton} in conjunction with {\em NuSTAR} observations, preferably when the $\gamma$-ray activity of the source is low, will be needed to place stronger constraints on the models.~A~significant step forward in the study of the X-ray spectra of $\gamma$-ray-emitting NLSy1 will be obtained thanks to the next generation of X-ray satellites (i.e., Athena, XRISM, \mbox{eXTP; \citep[]{barcons17,tashiro18,intzand19}}).

X-ray variability has been observed in most of the $\gamma$-ray-emitting NLSy1, in~particular during high $\gamma$-ray activity periods (e.g., \citep[]{dammando13a, paliya14, dammando15b, dammando16b,dammando19}), although~not extreme and rapid as observed in radio-quiet NLSy1 (e.g., IRAS 13224$-$3809; \citep[]{boller97}). The~contemporaneous high activity level observed in X-rays and $\gamma$-rays suggests a co-spatiality of the emitting region, with~the same mechanism (i.e.,~IC~scattering) responsible {for the radiation emitted in those energy bands}, as~observed in several FSRQ (e.g., \citep[]{dammando12b,raiteri12,wehrle12}).

At hard X-rays only one $\gamma$-ray-emitting NLSy1 is included in the {\em Swift}-BAT 105-month catalogue~\citep{oh18}, 1H 0323$+$342, with~a photon index of 1.62 $\pm$ 0.30. The~source is included also in the Fourth IBIS/ISGRI Soft Gamma-ray Survey Catalog \citep{bird10}. Moreover, PMN J0948$+$0022 is included in the preliminary 100-month {\em Swift}-BAT Palermo catalog\footnote{\url{http://bat.ifc.inaf.it/100m\_bat\_catalog/100m\_bat\_catalog\_v0.0.htm}.}. Both sources are detected in hard X-rays by {\em NuSTAR} during pointing observation \citep{landt17, brenneman17}, together with PKS 2004$-$447 \citep{dammando17}.

\section{Infrared and Optical~Properties}

Different components (star formation activity, relativistic jet, and~dusty torus) can contribute to the infrared emission of radio-loud NLSy1. Thanks to the all sky survey carried out by the {\em Wide-field Infrared Survey (WISE)} satellite a dedicated study of the infrared properties of a sample of 42 radio-loud NLSy1 has been performed by \citep{caccianiga15}, confirming that infrared colours can be reproduced by a combination of AGN (dusty torus and relativistic jet) and host galaxy emission (star-formation emission). In~particular, the~infrared colours of $\gamma$-ray-emitting NLSy1 in that sample (i.e., 1H 0323$+$342, SBS 0846$+$513, PMN J0948$+$0022, PKS 1502$+$036, FBQS J1644$+$2619, and~PKS 2004$-$447) {have shown} properties similar to FSRQ, with~a dominance of the jet emission component in the infrared band \citep{caccianiga15}.

By investigating {\em WISE} data, rapid infrared variability {has been} detected in some $\gamma$-ray-emitting NLSy1. In~particular, \citep{jiang12} found infrared intraday variability in SBS 0846$+$513 and PMN J0948$+$0022, with~an amplitude of $\sim$0.1--0.2 mag and an emitting region size obtained by causality argument as $<$ 10$^{-3}$ pc, significantly smaller than the scale of the torus but consistent with an origin from the jet base. A~similar variability amplitude was observed on a monthly time-scale for PKS 1502$+$036. Rapid~infrared variability over a time-scale consistent with the jet-emitting region has been observed also in TXS 2116$-$077 by WISE \citep{yang18}.

Optical intraday variability has been reported for PMN J0948$+$0022 \citep{liu10, maune13, itoh13, paliya13}, \mbox{SBS~0846$+$513~\citep{maune14, paliya16b}}, PKS 1502$+$036 \citep{paliya13}, and~1H 0323$+$342 \citep{itoh14, ojha19}\footnote{Optical intraday variability in this source has been reported also in \citep{paliya14}. However, the~contamination of the host galaxy has not be considered in the aperture photometry analysis used for producing the light curve of the source, making the results~doubtful.}. On~the other hand, no~intraday variability has been observed in optical for PKS 1502$+$036 \citep{ojha19}.

For PMN J0948$+$0022, optical polarization as high as 36$\%$ \citep{itoh13} has been observed, with~minute time-scale polarization variability but no variation of the electric vector position angle (EVPA). This~has been interpreted as evidence of synchrotron emission from a compact region of highly ordered magnetic field. Polarization as high as 10$\%$ and intraday variation have been observed also in SBS 0846$+$513, accompanied by a rapid change of the EVPA \citep{maune14}.

During an optical polarimetry monitoring of 1H 0323$+$342 after the high $\gamma$-ray activity state observed by {\em Fermi}-LAT in July 2013 and lasted $\sim$20 days, the~polarization of the source increased only to $\sim$3$\%$ \citep{itoh14}. A~similar low polarization value has been observed by \citep{angelakis18}. The~variable polarized emission has confirmed the synchrotron origin for the optical emission, although~the low optical polarization level of the source is probably due to the contamination from the thermal emission from the accretion disc. The~optical EVPA was parallel to the jet orientation, implying a magnetic field direction, associated with the optical emission, perpendicular to the jet axis \citep{itoh14}. This implies either a presence of helical/toroidal magnetic field or a magnetic field compressed by transverse shock propagating down the jet. The~different behaviours observed in these sources should be investigated in more details. However, the~lack of high cadence optical polarisation observations of radio-loud NLSy1 limits the possibility to perform a systematic population~study.

 Periods of significant polarisation amplitude and angle variability have been reported in \citep{angelakis18} for the $\gamma$-ray-emitting NLSy1 1H 0323$+$342, SBS 0846$+$513, PMN J0948$+$0022, and~PKS 1502$+$036. Moreover, the~EVPA of 1H 0323$+$342 has shown a preferential orientation (at 49.3 degrees to the position angle of the 15 GHz radio jet) similar to what was observed in high-synchrotron-peaked blazars \citep{angelakis16}. The~projected magnetic field of 1H 0323$+$342 is oriented at 40.7 degrees to the jet axis. This misalignment can be due to a combination of poloidal and toroidal field~components.

Three long rotations of the polarisation angle have been observed in PKS 1502$+$036 and 1H 0323$+$342.~Although~the observed behaviour can be induced by noise, an~intrinsic evolution of the EVPA is the most likely explanation of the observed behaviour. Rotations of the optical polarization plane have been observed in blazars, most of the time associated with high $\gamma$-ray activity \mbox{(e.g.,~\citep[]{marscher08, abdo10c, blinov16, blinov18})}.~This behaviour has been explained by different theoretical scenarios: rotation of the emitting region on a helical trajectory (e.g., \citep[]{marscher08}), propagation in large-scale bent jet (e.g.,~\citep[]{abdo10c}), turbulent processes resulting in random walks (e.g., \citep[]{marscher14}), light travel-time and projection effects within an axisymmetric emission region (e.g., \citep[]{zhang14}). Linear polarization observations of well sampled objects suggested that the relativistic jet in blazars contains a superposition of ordered helical magnetic fields and turbulent magnetic fields. In~fact, high degree of polarization is expected in case of synchrotron radiation from a region of uniform magnetic field. The~relatively low value of degree of optical polarization observed in blazars indicates that there is a mechanism in place that is the cause of disorder in magnetic field, like the turbulence. On~the other hand, the~constancy of EVPA in some periods supports the action of a process that partially orders the magnetic field, like a shock or a helical magnetic field pattern, superposed to the turbulence. The~observed patterns can be connected to the different contribution of the ordered and turbulent magnetic fields and the presence of shocks or not. These observed patterns can be compared to simulated light curves produced by the Turbulent Extreme Multi-Zone model to describe the time-dependent polarization of blazars (e.g.,~\citep[]{marscher17}). Long-term monitoring campaigns with more accurate observations are needed to compare the observed polarizarion data to simulated light curves in $\gamma$-ray-emitting NLSy1 and {to provide} clear evidence of similarity with blazars in their optical polarization~variability.

The optical and UV part of the spectrum of FSRQ is usually characterized by the thermal emission from an accretion disc, while in BL Lac objects the disc is radiatively inefficient, with~no significant emission in this energy range (e.g., \citep[]{ghisellini11}). The~accretion disc emission is visible in the low activity state of the SED of PMN J0948$+$0022 \citep{dammando15b}, 1H 0323$+$342 \citep{abdo2009b}, FBQS J1644$+$2619 \citep{larsson18}, TXS 2116$-$077~\citep{yang18}, and~PKS 1502$+$036 \citep{dammando16b}. On~the contrary, no significant evidence of thermal emission from the accretion disc has been observed in SBS 0846$+$513 \citep{dammando13a} and PKS 2004$-$447 \citep{orienti15}.

\begin{figure}[H]
\centering
\rotatebox{0}{\resizebox{!}{70mm}{\includegraphics{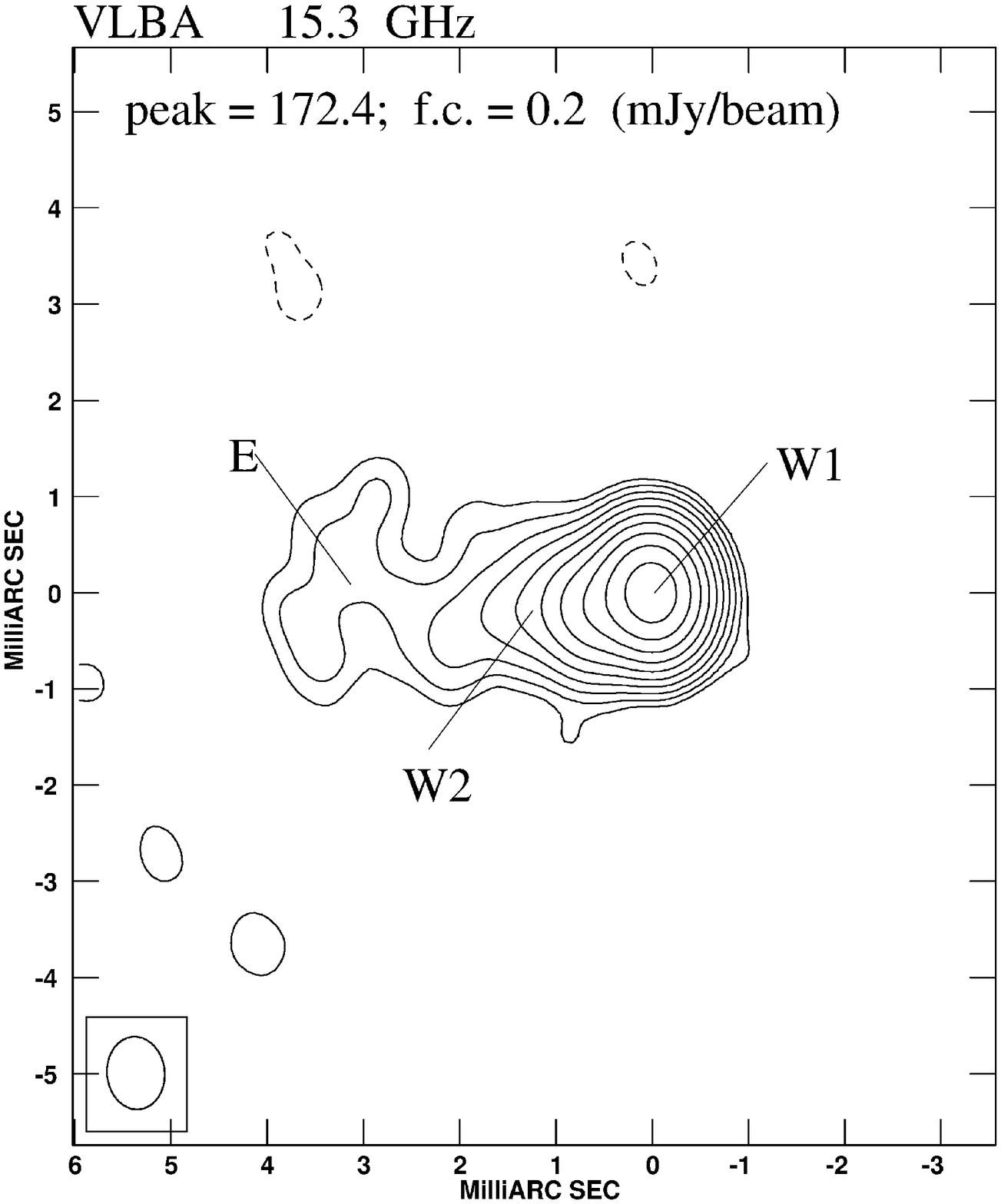}}}
\hspace{0.2cm}
\rotatebox{0}{\resizebox{!}{70mm}{\includegraphics{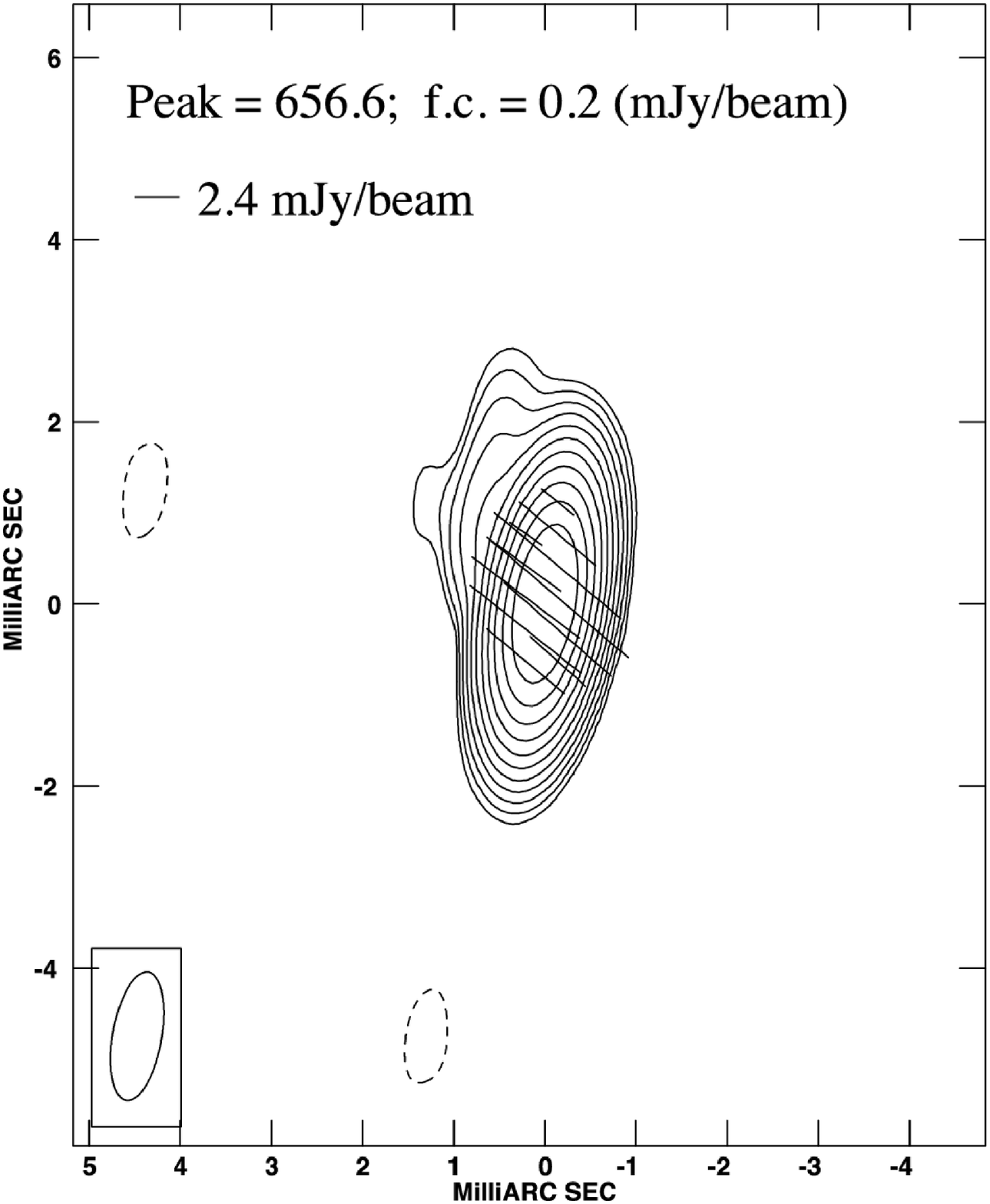}}}
\caption{{\it Left panel}: VLBA image at 15 GHz of SBS 0846$+$513. Component W1 corresponds to the core region, component W2 to a knot in the jet, component E is an extended low-surface brightness structure with a steep spectrum. Adapted from \citep{dammando12}. {\it Right panel}: VLBA image at 15 GHz of PMN J0948+0022 collected on 2011 May 26. The~vectors superimposed on the total intensity contours show the percentage polarization and the position angle of the electric vector. Adapted from \citep{dammando14}. In~both images, it is provided the restoring beam, plotted in the bottom-left corner, the~peak flux density in mJy/beam and the first contour (f.c.) intensity in mJy/beam, which is three times the off-source noise level. Contour levels increase by a factor of~2.}
\label{MOJAVE}
\end{figure}

\section{Radio~Properties}\label{sec5}

On pc scale a core-jet structure has observed for SBS 0846+513 \citep{dammando12}  (Figure~\ref{MOJAVE}, {left} panel), PKS~2004$-$447 \citep{orienti12, orienti15}, 1H 0323$+$342 \citep{wajima14,hada18}, FBQS J1644$+$2619 \citep{doi07,doi11}, B3 1441$+$476 \citep{gu15}, PKS~1502$+$036 \citep{dammando13b}, and~PMN J0948$+$0022 (Figure~\ref{MOJAVE}, {right} panel) \citep{giroletti11, dammando14}, although~the jet in the two latter sources is significantly fainter than that observed in the other sources. The~pc scale morphology of $\gamma$-ray-emitting NLSy1 seems to be a reminiscent of blazars, suggesting Doppler boosting effect due {to} small jet viewing angles. Two-sided radio morphology has been seen at kpc scale in 1H 0323$+$342, PMN J0948$+$0022, and~FBQS J1644$+$2619 \citep{anton08, doi12, doi19}. This challenges the possible connection between compact steep spectrum (CSS) radio sources and radio-loud NLSy1, in~particular $\gamma$-ray-emitting NLSy1 being the aligned population of NLSy1 where the CSS properties are hidden by boosting effects (e.g.,~\citep[]{oshlack01, komossa06, yuan08, caccianiga14, orienti15, gu15}). In~the same way the X-ray spectrum of CSS sources is typically highly obscured with column density higher than 10$^{22}$ cm$^{-2}$ (e.g., \citep[]{tengstrand09}). The~lack of evidence of intrinsic absorption in addition to the Galactic one in X-rays (see Section~\ref{Xray_properties}) disfavors a link between $\gamma$-ray-emitting NLSy1 and CSS~sources.

\begin{figure}[H]
\centering
\rotatebox{0}{\resizebox{!}{70mm}{\includegraphics{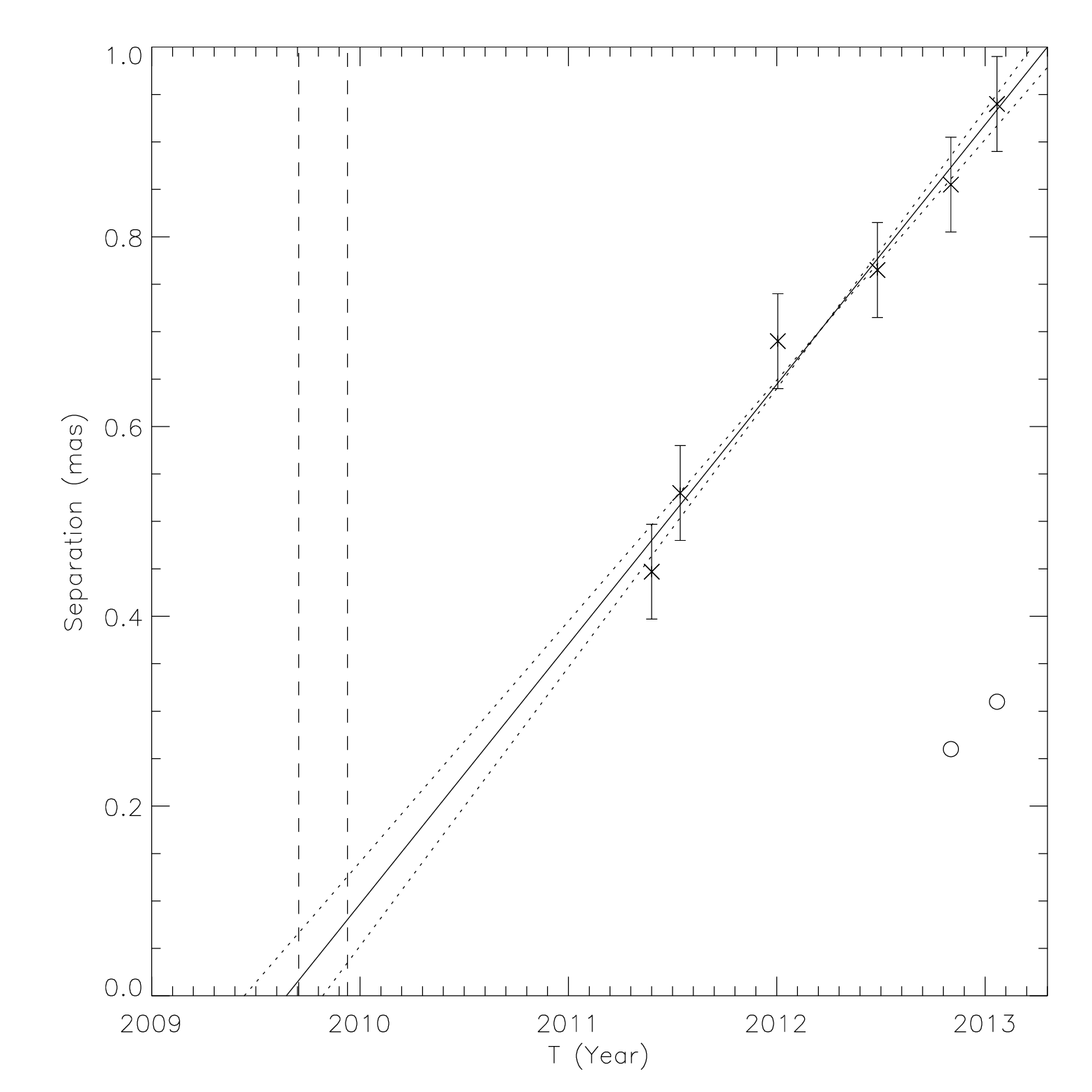}}}
\hspace{0.2cm}
\rotatebox{0}{\resizebox{!}{70mm}{\includegraphics{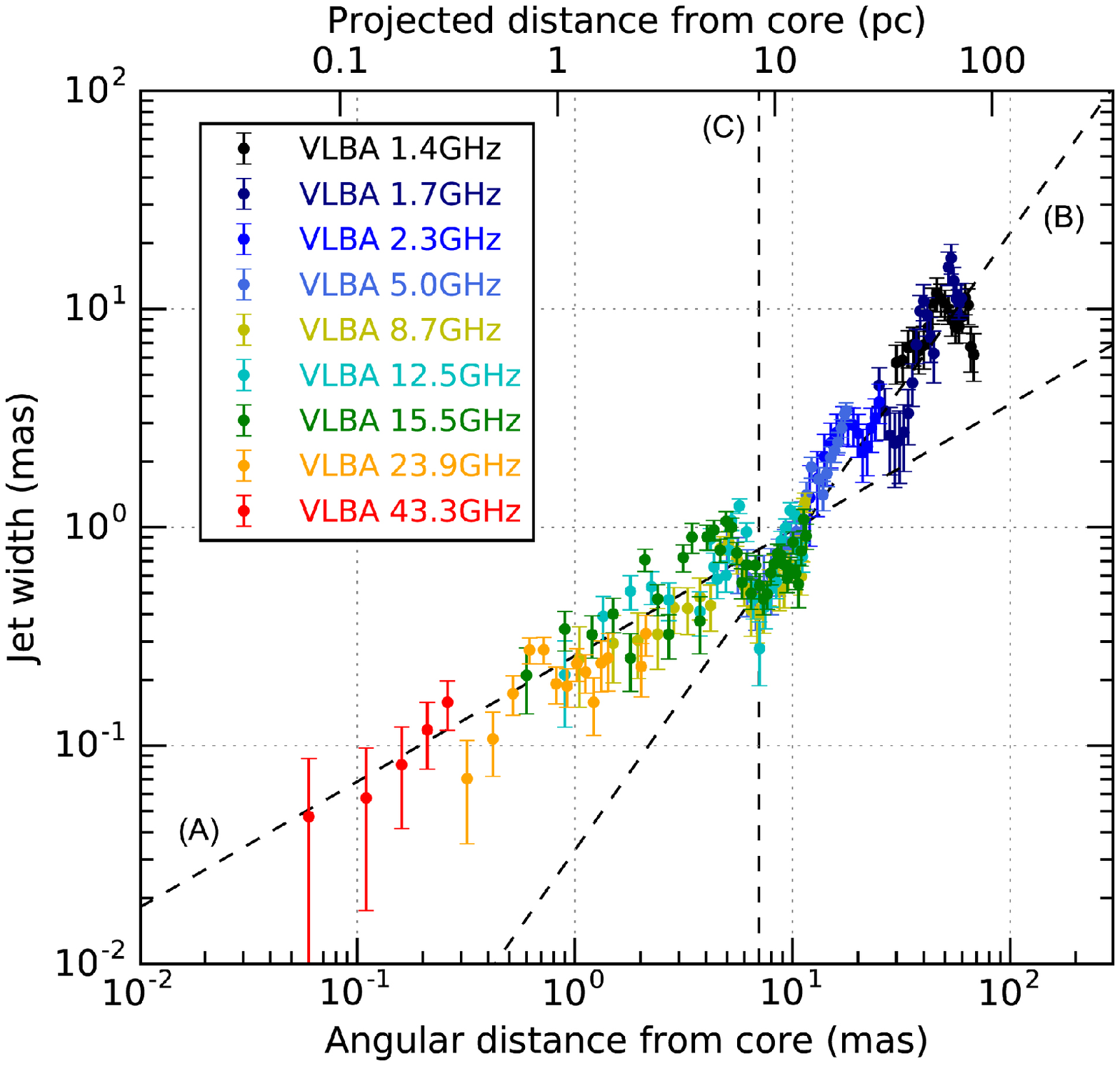}}}
\caption{{\it Left panel}: The separation between the core component of SBS 0846$+$513 and the knot ejected in 2009 as a function of time. The~solid line represents the regression fit to the 15 GHz VLBA MOJAVE data, while the dotted lines represent the uncertainties from the fit parameters. Dashed lines indicate the beginning and the peak of the radio flare observed by OVRO in 2009. Adapted~from~\citep{dammando13a}.  {\it Right~panel}: Jet~width profile of 1H 0323$+$342 as a function of (projected) distance $z$ from the 43~GHz core.~The~dashed line (A) represents a best-fit model for the inner jet ($z$ $<$ 6.5 mas), indicating a parabolic collimating jet. The~dashed line (B) represents a best-fit model for the outer ($z$ $>$ 7.5 mas) jet, indicating a hyperbolic expanding jet. The~vertical dashed line (C) indicates the location of the quasi-stationary feature S. Adapted from \citep{hada18}.}
\label{MOJAVE_VLBA}
\end{figure}

The analysis of multi-epoch Very Long Baseline Interferometry (VLBI) data of SBS 0846$+$513 collected by the MOJAVE programme\footnote{\url{https://www.physics.purdue.edu/MOJAVE/}.} during 2011--2013 pointed out a superluminal jet component moving away from the core with an apparent velocity of (9.3 $\pm$ 0.6)$c$ \citep{dammando13a} (Figure~\ref{MOJAVE_VLBA}, left panel), in~agreement with the results presented in \citep{lister16}. This apparent superluminal velocity supports the presence of boosting effects for the jet of SBS 0846$+$513. Interestingly, the~superluminal jet component was ejected close in time with a radio flare observed in December 2009 by the Owens Valley Radio Observatory (OVRO) but during a quiescent activity state in $\gamma$-rays, representing an example of knot emission not connected to any $\gamma$-ray flaring activity. On~the other hand, a~new jet component for SBS 0846$+$513 has been ejected close in time with the $\gamma$-ray flare occurred in June--July 2011. Highly superluminal features with an apparent velocity of $\sim$9$c$ have been reported also for PMN J0948$+$0022 and 1H 0323$+$342 \citep{lister16, doi18, lister19}.
On the contrary, no apparent superluminal motion {has been} detected for PKS 1502+036 during 2008--2012, although~the radio spectral variability, the~one-sided jet-like structure, and~the $\gamma$-ray properties seem to require the presence of boosting effects \citep{dammando13b}. The~lack of significant proper motion in PKS 1502$+$036 has been confirmed by \citep{lister16}, with~only a sub-luminal component detected. This result seems to resemble the `Doppler factor crisis’ observed in bright TeV BL Lacs (e.g., \citep[]{piner04, piner10}). However, the~SED of PKS 1502+036, in~particular the high Compton dominance\footnote{Compton dominance is the ratio of the peak Compton luminosity to peak synchrotron luminosity.} (see Section~\ref{SED}) does not resemble a TeV BL Lac, but~an FSRQ. This may suggest the presence of a structured jet in this NLSy1, with~different regions {having} different Lorentz factors (i.e., a~fast spine and a slower layer) (see e.g., \citep[]{piner18}). Unfortunately, no Very Long Baseline Array (VLBA) monitoring was available during the $\gamma$-ray flaring activity of PKS 1502$+$036 observed in 2016~\citep{dammando16b} to investigate the possible ejection of a new superluminal jet component during high $\gamma$-ray~states.

A low fractional polarization compared to BL Lacs and FSRQ has been obtained by the analysis of the core linear polarization properties of 4 $\gamma$-ray-emitting NLSy1 (i.e., 1H 0323$+$342, SBS 0846$+$513, PMN J0948$+$0022, and~PKS 1502$+$036) using 15 GHz VLBA data. In~addition, large misalignment between the core EVPA direction and the inner jet position angle has been observed in these 4 sources, similarly to what was observed in FSRQ. This can be related to a similarity with FSRQ in shock strength and geometry, with~a standing transverse shock at the base of the jet which collimates the jet magnetic field perpendicular to the jet direction \citep{hodge18}. For~1H 0323$+$342, the~nearest $\gamma$-ray-emitting NLSy1, the~morphology of the inner jet obtained by using high-resolution multifrequency VLBA observations is well characterized by a parabolic shape up to 7 mas, an~indication that the jet is continously collimated near the base of the jet (Figure~\ref{MOJAVE_VLBA}, right panel). Beyond~7 milliarcsecond (mas) the jet expands more rapidly at larger scales, transitioning into a hyperbolic/conical shape~\citep{hada18}. The~acceleration and collimation zone seem to co-exist in this source, ending up with a bright quasi-stationary feature at 7~mas (a~deprojected distance of $\sim$100 pc) corresponding to a recollimation shock \citep{doi18}. The~recollimation shock may be a possible site of $\gamma$-ray emission, similar to what is observed with HST-1 for the radio galaxy M87 (e.g., \citep[]{cheung07}). This may be an indication of a common jet formation mechanism for the two sources. The~location of the $\gamma$-ray {emitting region} on a standing shock located far away from the SMBH has been proposed also for some BL Lacs (e.g., \citep[]{marscher08,agudo11}) and FSRQ (e.g., \citep[]{marscher10,orienti13}).

Strong radio variability {has been} observed at 15 GHz during the monitoring of the OVRO 40-m telescope of PMN J0948+0022 \citep{dammando14,dammando15b}, PKS 1502+036 \citep{dammando13b}, and~SBS 0846+513 \citep{dammando12,dammando13a}. An~inferred variability brightness temperature of 2.5 $\times$ 10$^{13}$, 1.1 $\times$ 10$^{14}$, and~3.4 $\times$ 10$^{11}$~K is obtained for PKS 1502$+$036, SBS 0846$+$513, and~PMN J0948$+$0022, respectively. A~variability brightness temperature of 7.0 $\times$ 10$^{11}$~K {has been} obtained for 1H 0323$+$342 during the monitoring by the Yamaguchi 32-m radio telescope~\citep{wajima14}. In~case of PKS 2004$-$447 a variability brightness temperature of 1.7 $\times$ 10$^{14}$~K has been derived by \citep{orienti15} by using VLBA and Very Large Array (VLA) observations, in~agreement with the results obtained based on the Ceduna monitoring \citep{gallo06}.
These values are larger than the brightness temperature derived for the Compton catastrophe \citep{readhead94}, suggesting that the radio emission of the jet is Doppler boosted. On~the other hand, a~high variability brightness temperature of 10$^{13}$ K, comparable to that of $\gamma$-ray-emitting NLSy1, has been observed for TXS 1546$+$353. However, no $\gamma$-ray emission has been detected from that source so far \citep{orienti15}. The~brightness temperature derived from the 1.7 GHz VLBI flux density of the unresolved core of FBQS J1644$+$2619 (7 $\times$ 10$^{9}$~K) is too high for free-free emission, indicating that non-thermal processes dominate the radio emission, but~lower than the limit value predicted by the Compton catastrophe \citep{dammando15a}.

An intensive monitoring of the $\gamma$-ray-emitting NLSy1 1H 0323$+$342, SBS 0846$+$513, PMN~J0948$+$0022, and~PKS 1502$+$036 from 2.6 GHz to 142 GHz with the Effelsberg 100-m and IRAM 30-m telescopes showed, in~addition to an intensive variability, spectral evolution across the different bands following evolutionary paths which can be explained by shocks operating in a plasma outflow. These are typical characteristics seen in blazars \citep{angelakis15}.

\section{SED~Modelling}\label{SED}

\begin{figure}[H]
\centering
\includegraphics[width=120mm]{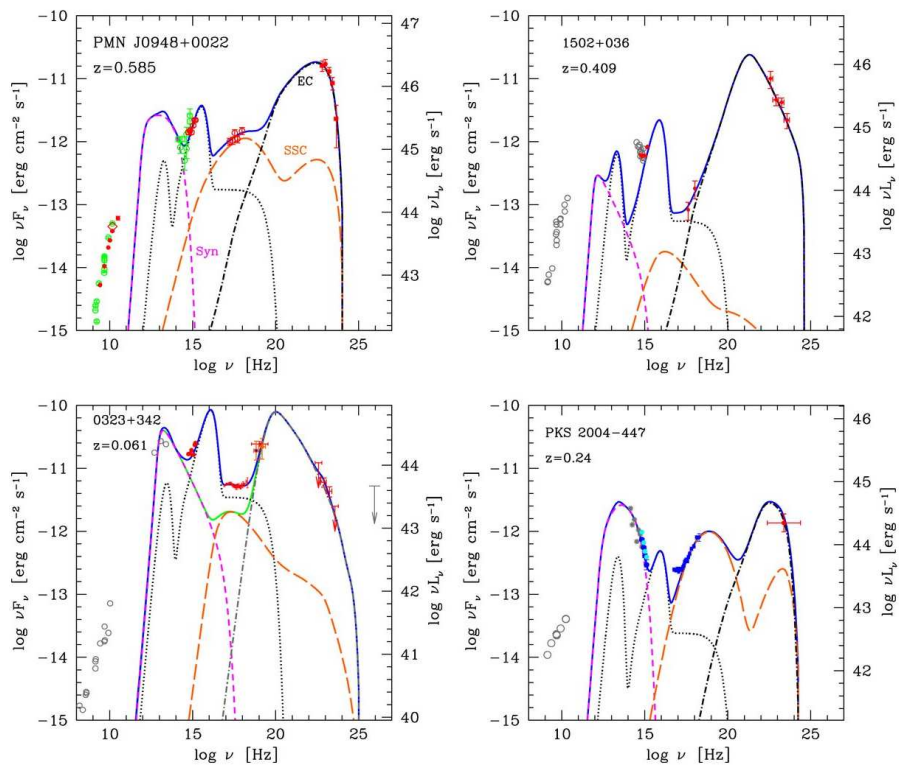}
\caption{SED of the four NLSy1 detected by {\em Fermi}-LAT during the first year of operation. The~synchrotron self-absorption is clearly visible around 10$^{11-12}$ Hz. The~short dashed light blue line indicates the synchrotron component, while the long dashed orange line is the synchrotron-self Compton emission. The~dot-dashed line refers to external Compton emission and the dotted black line represents the contribution of the accretion disc, X-ray corona and the infrared torus. The~continuous blue line is the sum of all the contributions. Adapted from \citep{abdo2009b}.}
\label{SED_NLSy1}
\end{figure}

The first SED collected for the four NLSy1 detected in the first year of {\em Fermi}-LAT operation showed clear similarities with blazars: a double-humped shape with a first peak in the infrared/optical band due to synchrotron emission, a~second peak in the MeV/GeV band likely due to IC emission, and~an additional component related to the accretion disc in UV for three of the four sources (Figure~\ref{SED_NLSy1}). The~physical parameters of these NLSy1 are blazar-like, and~the jet powers are in the range of values usually observed in blazars \citep{abdo2009b}. Further studies confirmed that the physical parameters of $\gamma$-ray-emitting NLSy1 are blazar-like, in~particular they seems to resemble FSRQ at the low-end of their BH mass distribution, with~the dominant mechanism for producing the $\gamma$-ray emission being the IC scattering of seed photons from BLR or the dusty torus and the jet power being {lower than} the values estimated for FSRQ (e.g., \citep[]{foschini11,dammando13a,dammando14,dammando15b,dammando16b}).

\begin{figure}[H]
\centering
\includegraphics[width=75mm]{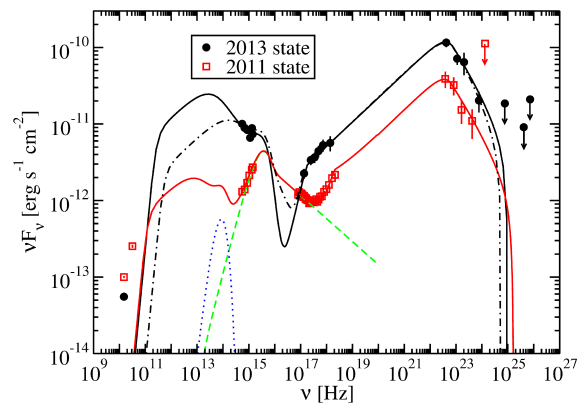}
\caption{SED and models for the 2013 and 2011 activity states of PMN J0948$+$0022. The~filled circles are the data from the 2013 flaring state, and~the open squares are the data from the 2011 intermediate state taken from \citep{dammando14}. The~dashed curve indicates the disc and coronal emission, and~the dotted line indicates the thermal dust emission. Solid lines represent models consistent with scattering dust torus radiation, while the dashed-dotted curve represents a model consistent with the scattering of BLR radiation. Arrows refer to 2$\sigma$ upper limits on the source flux. The~VERITAS upper limits are corrected for EBL absorption using the model of \citep{finke10}. Adapted from \citep{dammando15b}.}
\label{SED_0948}
\end{figure}

For PMN J0948$+$0022 the broad-band SED of the 2013 flaring activity state has been compared with that from an intermediate activity state observed in 2011 in \citep{dammando14} (Figure~\ref{SED_0948}). Contrary to what was observed for some FSRQ (e.g., PKS 0537$-$441; \citep[]{dammando13c}) the SED of the two activity states, modelled as synchrotron emission and external Compton scattering of seed photons from a dust torus, could not be modelled by changing only the electron distribution parameters. A~higher magnetic field is needed for the high activity state, consistent with the modelling of different activity states of the FSRQ PKS 0208$-$512 \citep{chatterjee13}. The~2013 flaring state has been modelled also assuming external Compton scattering of BLR photons. The~model reproduces the data as well as the scattering of the IR torus photons, but~requires magnetic fields which are far from equipartition \citep{dammando15b}.~On~the other hand, external Compton scattering of BLR photons has been proposed as main mechanism for producing the high-energy emission in case of the first $\gamma$-ray flare observed by the same source in July 2010 \citep{foschini11}. In~the same way, the~$\gamma$-ray emission from 1H 0323$+$342 has been modelled with external Compton scattering of BLR photons. For~this source a significant contribution from the X-ray corona has been taken into account to describe the X-ray part of the SED, at~least during some periods \citep{paliya14,kynoch18}.

\begin{figure}[H]
\centering
\rotatebox{0}{\resizebox{!}{55mm}{\includegraphics{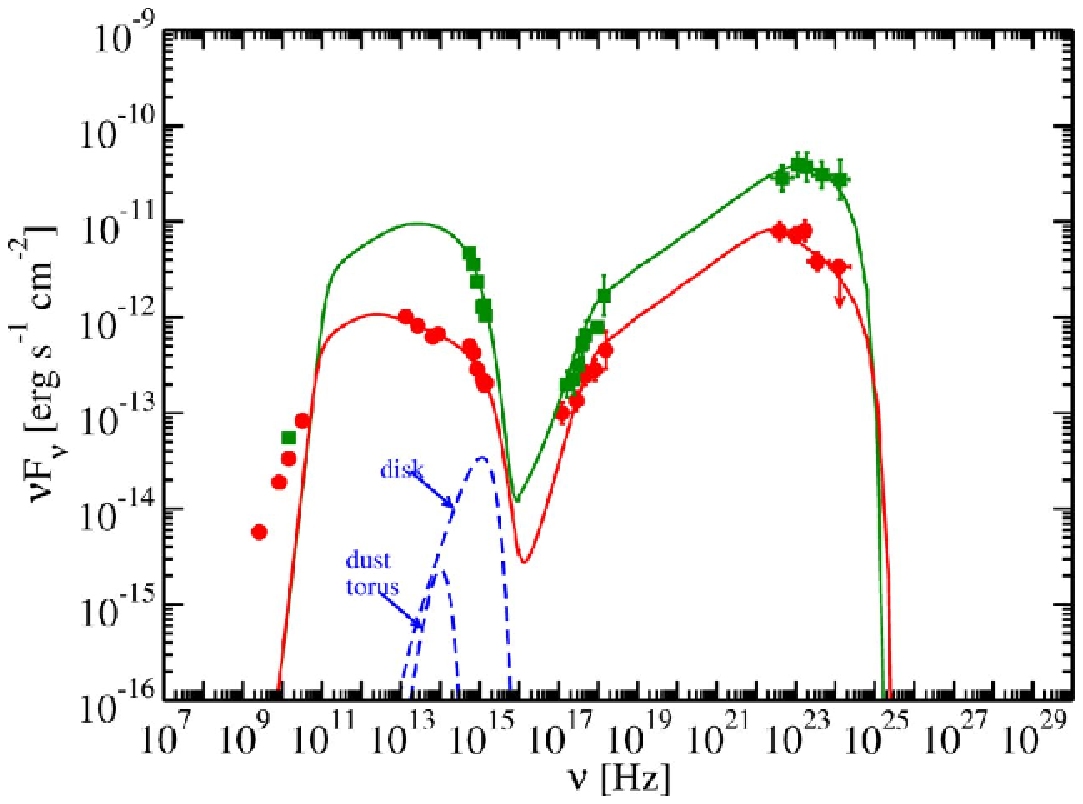}}}
\hspace{0.1cm}
\rotatebox{0}{\resizebox{!}{55mm}{\includegraphics{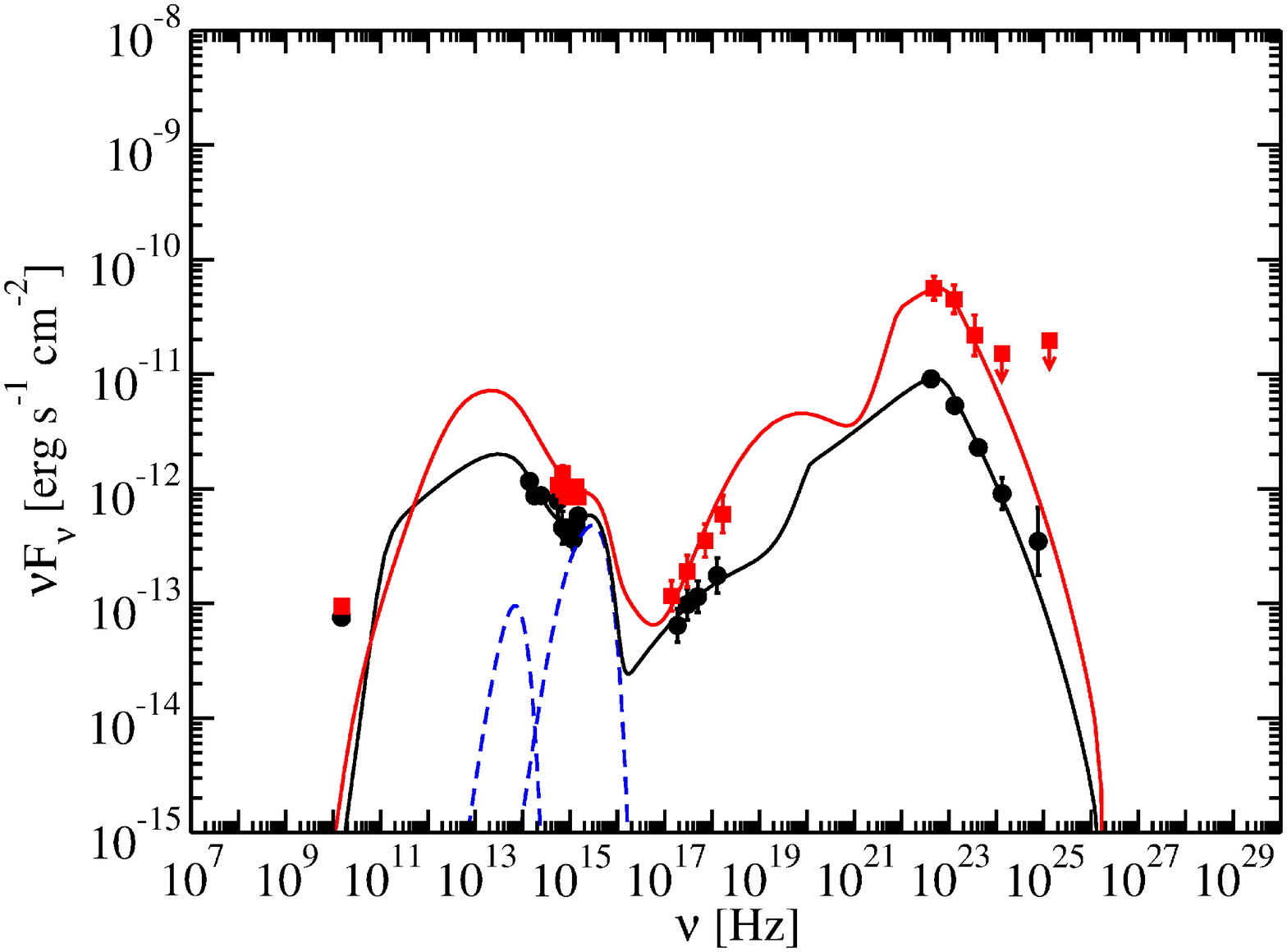}}}
\caption{SED data (squares) and model fit (solid curve) of SBS 0846$+$513 ({\it left panel}) and PKS 1502$+$036 ({\it right panel}) in flaring activity with the thermal emission components shown as dashed curves. The~data points were collected by OVRO at 15 GHz, {\em Swift} (UVOT and XRT) and {\em Fermi}-LAT. The~SED in the quiescent state (taken from \citep{dammando12}) and average state, respectively, ares shown as circles. Adapted~from~\citep{dammando13b} ({\it left panel}) and \citep{dammando16b} ({\it right panel}).} \label{SED_0846_1502}
\end{figure}

The comparison of the SED of SBS 0846$+$513 during the flaring state in May 2012 with the SED built during a quiescent state is shown in the left panel of Figure~\ref{SED_0846_1502}.~Similar to the case of PMN J0948$+$0022, the~SED of the two different activity states, modelled by external Compton scattering of seed photons from a dust torus, could be fitted by changing the electron distribution parameters as well as the magnetic field \citep{dammando13b}. A~significant shift of the synchrotron peak to higher frequencies was observed during May 2012 flaring episode, similar to what has been observed in some FSRQ (e.g.,~PKS 1510$-$089, CTA 102; \citep[]{dammando11,raiteri17}). On~the other hand, contrary to what was observed for PMN J0948$+$0022, no significant evidence of thermal emission from the accretion disc has been observed in SBS 0846$+$513.

\unskip

Two SED of PKS 1502$+$036 collected during an average activity state and the 2016 flaring state were compared in \citep{dammando16b} (Figure~\ref{SED_0846_1502}, right panel). Most of the optical data in the SED are explained by synchrotron emission (together with a contribution from an accretion disc with a luminosity of 6~$\times$~10$^{44}$ erg s$^{-1}$, a~value lower than the luminosity usually observed for the disc of FSRQ as well as of the $\gamma$-ray-emitting NLSy1 PMN J0948+0022), the~X-ray data by synchrotron self Compton emission, and~the high-energy bump is modelled by an external Compton component with seed photons from a dust torus. Unlike the SED modelling reported in \citep{paliya16a} for the same source, the~magnetic field and Doppler factor are not changed simultaneously between the average and high activity state models in order to modify the minimum number of parameters between activity states. Moreover, jet powers for both the average and high states of PKS 1502$+$036 in \citep{dammando16b} are near equipartition between the electron and magnetic field energy density, with~the electron energy density being slightly higher in both cases. The~two SED could be fitted by changing the electron distribution parameters as well as the magnetic field, similar to the case of the FSRQ PKS 2142$-$75 \citep{dutka13} and PKS 1424$-$418 \citep{buson14} as well as the NLSy1 PMN J0948$+$0022 and SBS 0846$+$513 discussed in \citep{dammando13b, dammando15b}. However, differently from those cases the magnetic field decreased during the flare with respect to the average state. The~two SED of PKS 1502$+$036 show a Compton dominance $\sim$10. A~clear correlation between Compton dominance and the rest-frame peak synchrotron frequency was reported in \citep{finke13} for blazars, related to the contribution of the external Compton component, which results to be higher for larger values of Compton dominance. The~high value observed in PKS 1502$+$036 indicates that the external Compton emission is the main mechanism for producing $\gamma$-rays, as~observed for several FSRQ (e.g., \citep[]{finke13}). This confirms the similarities between $\gamma$-ray-emitting NLSy1 and FSRQ. A~similar high Compton dominance has been observed in PMN J0948$+$0022 and SBS 0846$+$513 during low and high activity states \citep{dammando15b}.

As for SBS 0846$+$513, PMN J0948$+$0022, and~PKS 1502$+$036, the~high-energy bump of the SED of PKS 2004$-$447 is modelled with an external Compton scattering of dust torus seed photons in~\citep{orienti15}. The~disc luminosity of PKS 2004$-$447 obtained by the SED modelling is particularly weak, $L_{\rm\,disc}$~=~1.8~$\times$~10$^{42}$ erg s$^{-1}$, consistent with the value estimated on the basis of the optical spectrum~\citep{foschini15}. Evidence has
been building that IC on BLR photons is disfavored as the main $\gamma$-ray mechanism in FSRQ with respect to the IC on IR photons from the torus \citep{costamante18}, in~agreement with the SED modelling proposed for these $\gamma$-ray-emitting~NLSy1.

\section{Host Galaxy and BH~Mass}\label{host_mass}

The mechanisms for producing relativistic jets in radio-loud AGN are still unclear. In~particular, the~physical parameters that drive the jet formation are under debate. One of the key parameters should be the BH mass, with~only large masses allowing an efficient jet formation \citep{sikora07}. Powerful relativistic jets are apparently only associated with the most massive SMBH ($M_{\rm\,BH}$ $>$ 10$^{8}$\,M$_{\odot}$; e.g.,~\citep[]{chiaberge11}) hosted in elliptical galaxies. In~a framework in which the jet is powered by energy extracted from the rotating SMBH (e.g., \citep[]{ghisellini14}), that can be interpreted as an evidence that efficient jet formation requires rapidly spinning SMBH, which can be obtained through major mergers \citep{sikora07}. In~fact, higher spins are predicted due to spin-up for episodes of coherent accretion with a preferred orientation, as~in the case of major mergers, while lower spins are expected for growth via minor mergers, in~which the angular momentum is deposited from random directions. Indeed, major mergers are commonly observed in the elliptical galaxies producing the most powerful jets \citep{ramos11, chiaberge15}.

In this context the discovery of relativistic jets in radio-loud NLSy1 usually associated with {relatively small} BH masses challenges the current knowledge on how the jets are generated and developed (e.g., \citep[]{bottcher02}).~The~estimates of the BH masses based on the virial method  (i.e., on~their luminosity and the width of their broad lines; e.g., \citep[]{bentz15}) for radio-loud NLSy1 are typically 10$^6$--10$^7$~M$_{\odot}$~\citep[][]{yuan08, mathur12}. These estimates suggest two possible interpretations: either radio-loud NLSy1 correspond to a fundamentally different mechanism for the formation of relativistic jets or alternatively the BH mass in radio-loud NLSy1 is largely~underestimated.

The BH mass of radio-loud NLSy1 might be underestimated due either to the effect of radiation pressure on BLR clouds \citep{marconi08} or to projection effects which reduces the width of their broad lines (\mbox{e.g.,~\citep[]{baldi16,decarli08}}).~The~effect of flattening of the BLR on the virial mass estimate may be larger in $\gamma$-ray-emitting NLSy1 that should have small angle of view, as~suggested by their blazar-like behaviour. Higher BH masses are instead in agreement with the values estimated by modeling the optical and UV data with a Shakura and Sunyaev disc spectrum (e.g., \citep[]{calderone13}). A~possibility way to reduce the discrepancy between the BH mass values obtained with the virial method and those obtained by the disc modelling method is to assume a radiatively efficiency of the disc  significantly lower than the 10$\%$ value usually observed for blazars \citep{calderone18}.

For PKS 1502$+$036 a BH mass of $\sim$7 $\times$ 10$^{8}$\,M$_{\odot}$ has been obtained from the near-infrared bulge luminosity of the host galaxy \citep{dammando18}. From~its optical spectrum and the BLR radius--luminosity relation by \citep{kaspi05}, a~virial mass of 4 $\times$ 10$^6$\,M$_\odot$ has been estimated by \citep{yuan08}.~On~the other hand, by~modelling the optical--UV data with a Shakura and Sunyaev accretion disc spectrum, \citep{calderone13} found $M_{\rm BH}$~=~3~$\times$ 10$^{8}$\,M$_{\odot}$, compatible with the value obtained by the near-infrared bulge luminosity within the~uncertainties.

Conflicting results were also obtained for FBQS J1644$+$2619, with~a BH mass of 2.1 $\times$ 10$^8$\,M$_\odot$ obtained by the near-infrared bulge luminosity \citep{dammando17}, significantly larger than the virial estimate (0.8--1.4~$\times$~10$^7$\,M$_\odot$; \citep{yuan08,foschini15}), but~compatible with the value obtained from the modelling of the accretion disc emission (1.6 $\times$ 10$^{8}$\,M$_{\odot}$; \citep{calderone13}).
Similarly, for~1H~0323$+$342 values in the range (1.5--2.2) $\times$ 10$^7$\,M$_\odot$ were estimated from near-infrared and optical spectroscopy \citep{landt17}, while values of (1.6--4.0) $\times$ 10$^8$\,M$_\odot$ were obtained by the near-infrared bulge luminosity \citep{leon14}.

It appears that the BH masses estimated with the virial method in $\gamma$-ray-emitting NLSy1 are systematically and significantly smaller that those derived from other techniques. It is worth mentioning that fitting the optical-UV spectrum of a sample of both radio-quiet and radio-loud NLSy1 (including 4 of the 9 $\gamma$-ray-emitting NLSy1) with the standard Shakura $\&$ Sunyaev accretion disc, \citep{viswanath19} has obtained BH masses that are about an order of magnitude larger than their virial estimates.
A higher BH mass estimation may solve the problem of the minimum BH mass required for the formation of relativistic jets, but~it leaves open the host galaxy issue. Spiral galaxies are usually formed by secular processes, with~central BH masses typically ranging between 10$^{6}$--10$^{7}$ M$_\odot$ (e.g.,~\citep[]{pastorini07, atkinson05}), so it would not be clear how powerful relativistic jets could form in spiral galaxies. Host galaxy studies of NLSy1 have mostly concentrated on radio-quiet objects (e.g., \citep[]{crenshaw03, ohta07}): most of them are found in disc-like galaxies (e.g., \citep[]{deo06}). However, only a handful of radio-loud NLSy1 have been investigated so~far.

\begin{figure}[H]
\centering
\includegraphics[width=120mm]{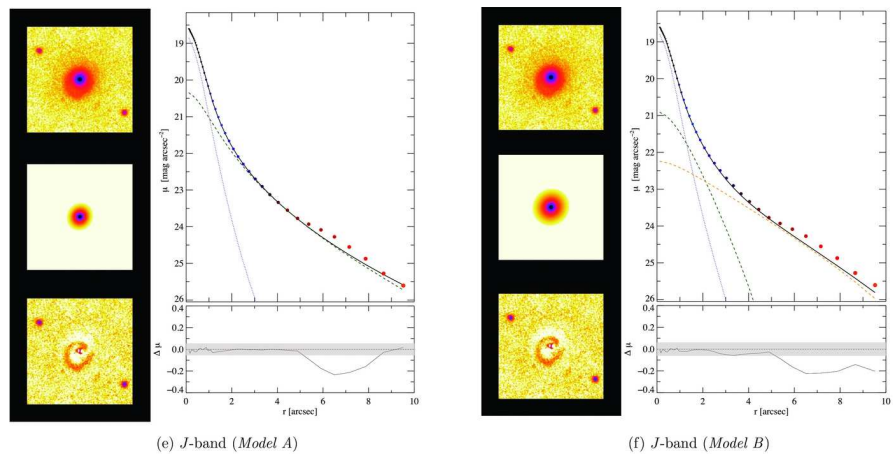}
\caption{Two-dimensional surface-brightness profile decomposition of 1H 0323$+$342 in $J$-band for Model A (PSF+Bulge; {\it left panel}) and Model B (PSF+Bulge+Disc; {\it right panel}). {\it Top left subpanel}: the~observed image in a field of view of 20 arcsec $\times$ 20 arcsec. {\it Middle left subpanel}: model used to describe the surface brightness distribution. {\it Bottom left subpanel}: the residual image. {\it Top right subpanel}: radial profile of the surface brightness distribution. The~filled circles show the observations, and~the solid, pointed, and~dashed lines represent the model, PSF, and~host galaxy, respectively. The~exponential disk component is shown in orange. {\it Bottom right subpanel}: residuals. Adapted from \citep{leon14}.}
\label{Host_0323}
\end{figure}

The morphology of the host galaxy has been determined only for 5 radio-loud NLSy1, four of them detected in $\gamma$-rays by {\em Fermi}-LAT. Observations of 1H 0323$+$342 with the Hubble Space Telescope and the Nordic Optical Telescope (NOT) revealed a structure that may be interpreted either as a one-armed galaxy \citep{zhou07} or as an elliptical galaxy with residual of a galaxy merger \citep{leon14} (Figure~\ref{Host_0323}). In~the case of PKS 2004$-$447, near-infrared observations with ISAAC mounted on Very Large Telescope (VLT) suggested that the host may have a pseudo-bulge morphology \citep{kotilainen16}. This should imply that the relativistic jet in PKS 2004$-$447 is launched from a pseudo-bulge via secular processes, in~contrast to the theoretical models proposed for the jet production. However, the~surface brightness distribution of the host is not well constrained by a bulge+disc model, leaving the debate on its morphology at large radii ($>$1.5--2 arcsec) still~open.

\begin{figure}[H]
\centering
\includegraphics[width=150mm]{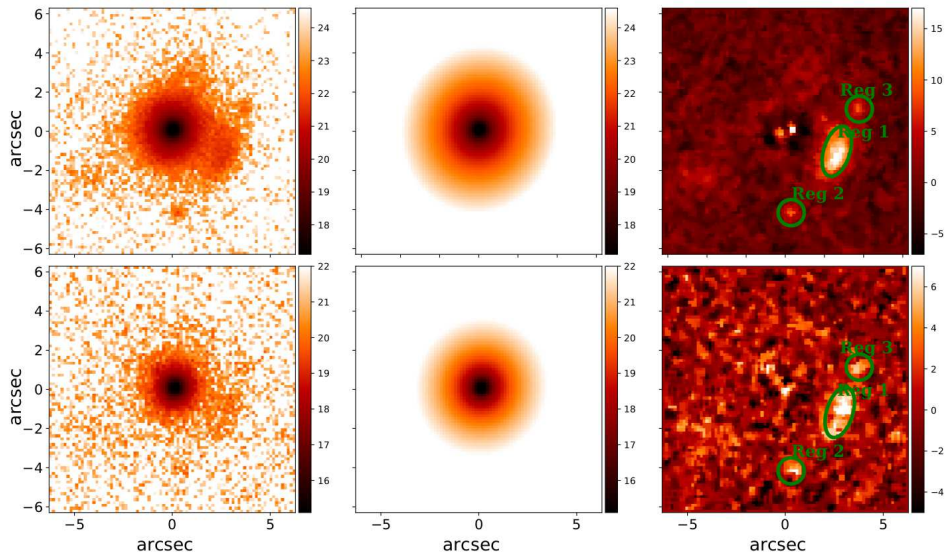}
\caption{{\it Left panels}: central 13 $\times$ 13\,arcsec$^{2}$ of the images in the $J$ and $K_s$ band of PKS 1502$+$036, top and bottom, respectively. {\it Center panels}: GALFIT models using a S\'ersic profile combined with a nuclear PSF. {\it Right panels}: residual images after subtracting the model. Colour bars are in mag\,arcsec$^{-2}$ (left-hand and centre panels). Adapted from \citep{dammando18}.}
\label{Host2}
\end{figure}

\begin{figure}[H]
\centering
\includegraphics[width=90mm]{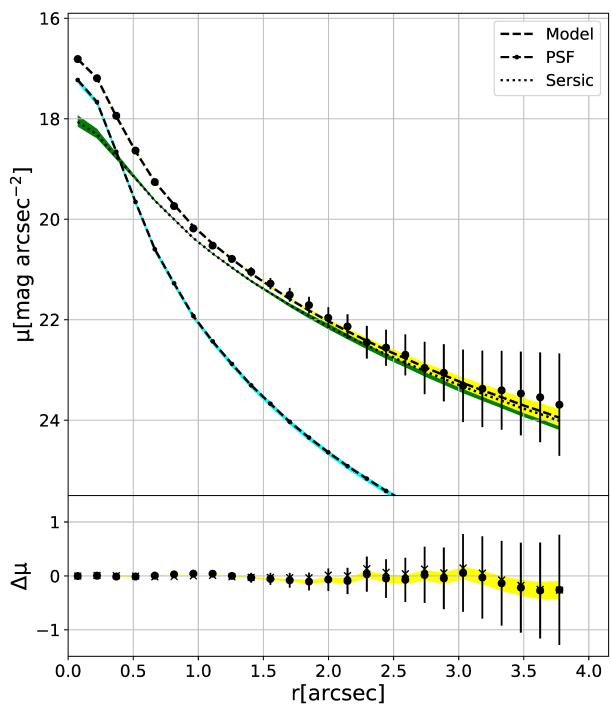}
\caption{Surface brightness decomposition in the $J$-band of PKS 1502$+$036. The~observed profiles are the black dots, the~nuclear PSF is the dot--dashed light blue curve, the~bulge component is reproduced with a S\'ersic model (green dot line). In~the bottom panels the residuals are shown. In~the residual panel crosses represent bulge + disc component. Adapted from \citep{dammando18}.}
\label{Host3}
\end{figure}

Observations of FBQS J1644$+$2619 in $J$ and $K_s$ bands with the NOT by \citep{olguin17}  suggested that the source resides in a barred lenticular galaxy. However, deeper near-infrared observations of FBQS J1644$+$2619 in $J$ band have been obtained using the Canarias InfraRed Camera Experiment (CIRCE) at the Gran Telescopio Canaries \citep{dammando17}. The~2D surface brightness profile of the source is modelled up to 5~arcsec by the combination of a nuclear and a bulge component with a S\'ersic index of 3.7, indicative of an elliptical galaxy. The~structural parameters of the host are consistent with the correlations of effective radius and surface brightness against absolute magnitude measured for elliptical galaxies. From~the infrared bulge luminosity \citep{marconi03} a BH mass of (2.1 $\pm$ 0.2) $\times$10$^{8}$ M$_\odot$ has been estimated. All~these pieces of evidence strongly indicate that the relativistic jet in FBQS J1644$+$2619 is produced by a massive SMBH in an elliptical galaxy, as~expected for radio-loud AGN. Galaxies residing in denser large-scale environments are preferably ellipticals (e.g., \citep[]{einasto14,kuutma17}). It is worth mentioning that FBQS J1644$+$2619 lies in a supercluster environment\footnote{Private communication by E. J\"arvel\"a.}, as~well as many radio-loud NLSy1, with~a trend of increasing radio loudness with increasing large-scale environment density. Under~the assumption that the large-scale environment may have an impact on the galaxy evolution, and~therefore the production of jet, the~dense large-scale environment of FBQS J1644$+$2619 is in agreement with the fact that the source is hosted in an elliptical galaxy \citep{jarvela17}.
The analysis of observations in $J$ and $K_s$ bands with the Infrared Spectrometer And Array Camera (ISAAC) on VLT of PKS 1502$+$036 (Figure~\ref{Host2}) has shown that its surface brightness profile, extending to $\sim$20 kpc, is modeled by the combination of a nuclear and a bulge component with a S\'ersic profile with index $n$ = 3.5, which is indicative of an elliptical galaxy (Figure~\ref{Host3}). From~the near-infrared bulge luminosity a BH mass of $\sim 7 \times 10^{8}$\,M$_{\odot}$ has been estimated. A~circumnuclear structure observed near PKS 1502$+$036 may be the result of a galaxy interaction \citep{dammando18}.

Near-infrared observations of the host galaxy of SDSS J161259.83$+$421940.3 (not detected in $\gamma$-rays so far) with NOT in $J$-band suggested a structure consistent with either a spiral arm or a morphological disturbance due to a recent merger, and~a BH mass of $\sim$8 $\times$ 10$^{6}$ $M_{\odot}$ \citep{jarvela18}. This is in agreement with the fact that $\gamma$-ray-emitting radio-loud NLSy1 (with a powerful relativistic jet) and radio-loud NLSy1 not able to emit in $\gamma$-rays (with a less powerful jet) can be part of different populations.
Tentative investigation of the host galaxy of SBS 0846$+$513 \citep{paliya18} and TXS 2116$-$077 \citep{yang18} by using SDSS images is not conclusive and strongly limited by the low resolution of the SDSS images for sources at redshift $z > 0.2$ .

For two of the nine $\gamma$-ray-emitting NLSy1 there is evidence that they reside in elliptical galaxies with a BH mass higher than 10$^{8}$ $M_{\odot}$, similar to blazars. A~possible scenario is that powerful relativistic jets can be produced only in radio-loud NLSy1 residing in elliptical galaxies. However, the~lack of information about the host galaxy for about half of the $\gamma$-ray-emitting NLSy1 together with some contrasting results indicate that the controversy on which galaxies host radio-loud NLSy1 is still open. In~addition, a~BH mass of 3--4 $\times$ 10$^{7}$ $M_{\odot}$ has been derived for 1H 0323$+$342 from reverberation mapping results obtained with the Lijiang 2.4-m telescope by \citep{wang16}. However, uncertainties on such a kind of measurements due to the effect of high accretion rate on the dynamics and geometry of the BLR in NLSy1 together with the adequacy of the cadence of the reverberation mapping campaign (e.g.,~\citep[]{czerny18}) and the viral factor used (e.g.,~\citep[]{yu19}) should be careful taken into account.
In case the low value of the BH mass estimated is confirmed, this would suggest that even low mass SMBH can produce powerful relativistic jets and represent a dramatic change in our view of radio-loud AGN. In~fact, the~connection between very massive SMBH and powerful radio emission is not purely an empirical threshold on the BH mass. It is becoming increasingly clear that radio-loud AGN are the result of a specific path of galaxies evolution, involving mergers, coalescence and spin-up of SMBH (e.g., \citep[]{capetti06, chiaberge15}). These mechanisms are expected to operate only at the high end of the galaxies mass distribution, and, consequently, only in the presence of very massive black holes. Only a handful of powerful radio galaxies have been found in spirals (e.g., \citep[]{morganti11, singh15}), all of them with BH mass higher than 10$^{8}$~M$_\odot$. Genuine low mass SMBH producing a radio-loud AGN would represent a strong challenge to this~picture.

\section{Conclusions}\label{sec8}

The discovery of $\gamma$-ray emission from radio-loud NLSy1 has raised important questions about the nature of these sources, the~conditions that lead to powerful jet formation and the mechanisms that produce the high-energy emission in these objects. Observations with the {\em Fermi Gamma-ray Space Telescope} have revealed NLSy1 as a possible new class of $\gamma$-ray-emitting AGN with blazar-like properties \citep{abdo2009b}. It is a small class, consisting of only 9 bona-fide NLSy1 to date \citep{abdollahi19}. The~continuous all-sky survey in $\gamma$-rays by {\em Fermi}-LAT will allow us to identify new $\gamma$-ray-emitting NLSy1 and to better characterize already known $\gamma$-ray sources belonging to this intriguing class of objects. However, the~search of transient $\gamma$-ray activity from NLSy1 by generating weekly and monthly light curve has not be effective yet \citep[e.g.,][]{abdollahi17}. In~addition, a~careful analysis of the current and future spectroscopical optical survey data is important to identify new bona-fide radio-loud NLSy1. In~this context, SDSS-V will be an all-sky multi-epoch spectroscopic survey of over six million objects \citep{kollmeier17} that allows an important step forward in the research and characterization of AGN classes including the~NLSy1.

The $\gamma$-ray observations as well as the multi-wavelength properties of $\gamma$-ray-emitting NLSy1 provide clear evidence for the existence of powerful jets pointed close to our line of sight in these objects. Compared to the blazar population, $\gamma$-ray-emitting NLSy1 seems to be similar to FSRQ, but~with lower jet powers. However, the~estimated BH mass of radio-loud NLSy1 has large uncertainties. These~uncertainties may influence the jet power and accretion rate estimates, and~therefore the comparison with $\gamma$-ray-emitting blazars. The~SED of $\gamma$-ray-emitting NLSy1 are Compton-dominated, with~the high-energy emission mainly produced by external Compton of infrared torus, resembling the typical SED of FSRQ. Differently from the $\gamma$-ray spectra, in~X-rays the spectra of $\gamma$-ray-emitting NLSy1 show not only the emission from the relativistic jet but also some Seyfert-like features from the accretion flow, like the soft X-ray excess and the Fe line in case of 1H 0323$+$342, unusual for blazars. A~regular monitoring from radio-to-$\gamma$-rays will be fundamental for continuing to investigate with great details the nature and emission mechanisms of these objects and to reveal further differences and similarities with~blazars.

Understanding the nature of the host galaxies of $\gamma$-ray-emitting NLSy1 and estimating their BH mass are of great interest in the context of the models for the formation of relativistic jets \citep[e.g.,][]{hopkins05}. There are increasing evidence that the host of some radio-loud NLSy1, in~particular $\gamma$-ray-emitting NLSy1, differ from those of radio-quiet NLSy1, usually spirals with low BH mass. In~case of FBQS J1644$+$2619 and PKS 1502$+$036 the host is an elliptical galaxy with a BH mass higher than 10$^{8}$~M$_\odot$, in~agreement to what is observed in radio-loud AGN. Estimates of the BH mass obtained with different techniques (i.e., accretion disc model fitting, optical spectro-polarimetry, IR bulge luminosity) are larger than the virial masses of these $\gamma$-ray-emitting NLSy1.~These results seem to confirm that a massive SMBH is a key ingredient for developing powerful relativistic jet and among the radio-loud NLSy1 only those hosted in massive elliptical galaxies are able to produce these structures. However, new~high-resolution observations of the host galaxies of other $\gamma$-ray-emitting NLSy1 will be fundamental to obtain further important insights into relativistic jet~development.

\vspace{6pt}



\authorcontributions{F.D. is the sole author of this~publication.}

\funding{This research received no external funding.}
\acknowledgments{I would like to thank all my collaborators that have worked with me on NLSy1 over the years, in~particular M. Orienti, J. Finke, C. M. Raiteri, J. Larsson, J. Acosta-Pulido, A. Capetti, M. Giroletti, T. Hovatta, E. Angelakis. This work was supported by the Korea’s National Research Council of Science and Technology (NST) granted by the International joint research project (EU-16-001). I acknowledge financial contribution from the agreement ASI-INAF n. 2017-14-H.0 and from the contract PRIN-SKA-CTA-INAF~2016.
This research has made use of NASA’s Astrophysics Data System. This research has made use of the NASA/IPAC Extragalactic Database, which is funded by the National Aeronautics and Space Administration and operated by the California Institute of~Technology.
The Fermi LAT Collaboration acknowledges generous ongoing support from a number of agencies and institutes that have supported both the development and the operation of the LAT as well as scientific data analysis. These include the National Aeronautics and Space Administration and the Department of Energy in the United States, the~Commissariat à l’Energie Atomique and the Centre National de la Recherche Scientifique/Institut National de Physique Nucléaire et de Physique des Particules in France, the~Agenzia Spaziale Italiana and the Istituto Nazionale di Fisica Nucleare in Italy, the~Ministry of Education, Culture, Sports, Science and Technology (MEXT), High Energy Accelerator Research Organization (KEK) and Japan Aerospace Exploration Agency (JAXA) in Japan, and~the K. A. Wallenberg Foundation, the~Swedish Research Council and the Swedish National Space Board in Sweden. Additional support for science analysis during the operations phase is gratefully acknowledged from the Istituto Nazionale di Astrofisica in Italy and the Centre National d’Études Spatiales in France. This work performed in part under DOE Contract DE-~AC02-76SF00515.
This research has made use of the data from the MOJAVE database that is maintained by the MOJAVE team (Lister~et~al. 2009, AJ, 137, 3718). Based~on observations obtained with XMM-Newton, an~ESA science mission with instruments and contributions directly funded by ESA Member States and NASA. The~National Radio Astronomy Observatory is a facility of the National Science Foundation operated under cooperative agreement by Associated Universities, Inc. I~acknowledge the use of public data from the Swift data archive. Based on observations made with ESO Telescopes at the La Silla Paranal Observatory under programme 290.B-5045. Based on observations made with the GTC telescope, in~the Spanish Observatorio del Roque de los Muchachos of the Instituto de Astrofisica de Canarias, under~Director’s Discretionary Time (proposal code GTC2016-053).}

\conflictsofinterest{The authors declare no conflict of~interest.}

\end{document}